\documentclass[%
 aip,
 amsmath,amssymb,
 reprint,%
]{revtex4-2}

\usepackage[]{graphicx}
\usepackage{subcaption}
\usepackage{dcolumn}
\usepackage{bm}
\usepackage[utf8]{inputenc}
\usepackage[T1]{fontenc}
\usepackage{mathptmx}
\usepackage{etoolbox}
\usepackage{comment}
\usepackage{siunitx}
\usepackage{hyperref}
\usepackage{float}

\graphicspath{{Img/}}

\makeatletter
\def\@email#1#2{%
 \endgroup
 \patchcmd{\titleblock@produce}
  {\frontmatter@RRAPformat}
  {\frontmatter@RRAPformat{\produce@RRAP{*#1\href{mailto:#2}{#2}}}\frontmatter@RRAPformat}
  {}{}
}
\makeatother

\begin{document}
\preprint{AIP/123-QED}
\title[]{A Pulsed Live-Cell Quantum Microscope for Entangled Solid State and Biological Qubits}

\author{Javier No\'e Ramos-Silva}
\altaffiliation{These authors contributed equally}
\affiliation{
Department of Electrical Engineering and Computer Science, University of California, Irvine, Irvine, CA 92697, USA
}
\author{Sangjun Noh }
\altaffiliation{These authors contributed equally}
\affiliation{
Department of Electrical Engineering and Computer Science, University of California, Irvine, Irvine, CA 92697, USA
}
\author{Minghao Jiang }
\affiliation{
Department of Physics and Astronomy, University of California, Irvine, Irvine, CA 92697, USA
}
\author{Parisa Aghaei }
\affiliation{
Department of Electrical Engineering and Computer Science, University of California, Irvine, Irvine, CA 92697, USA
}
\author{Ayla Hazrathosseini }
\affiliation{
Department of Physics, Texas A\&M University, College Station, TX 77843-3128, USA
}
\author{Jocelyn Leon }
\affiliation{
Department of Physics, Texas A\&M University, College Station, TX 77843-3128, USA
}
\author{Philip Hemmer }
\affiliation{
Department of Electrical and Computer Engineering, Texas A\&M University, College Station, TX 77843-3128, USA
}
\author{Peter J. Burke}
\email{pburke@uci.edu}
\affiliation{
Department of Electrical Engineering and Computer Science, University of California, Irvine, Irvine, CA 92697, USA
}
\affiliation{ 
Department of Biomedical Engineering, University of California, Irvine, CA 92697, USA
}
\affiliation{ 
Department of Chemical and Biomolecular Engineering, University of California, Irvine, CA 92697, USA
}
\affiliation{ 
Department of Materials Science and Engineering, University of California, Irvine, CA 92697, USA
}
\date{\today}

\begin{abstract}
Two revolutions in quantum sensing are converging on the same microscope stage. Biological qubits have emerged as genetically encoded, optically addressable quantum systems inside live cells. Solid state spin based qubits have been entangled and used as nanoscale correlator magnetometers, delivering sensitivity and spatial-resolution gains that single-spin probes cannot reach. Here we report a pulsed quantum microscope that enables simultaneous quantum state manipulation of both qubit technologies in live cells. The platform combines nanosecond-gated optical excitation at \SI{450}{\nano\meter} and \SI{520}{\nano\meter}, three-dimensional diffraction-limited addressing by galvo beam scanning and piezo objective focus, rapidly switched static and radio-frequency magnetic fields, single-photon timing with picosecond resolution, and microwave control for frequencies from DC to \SI{5}{\giga\hertz}, including the \SI{2.87}{\giga\hertz} resonance of solid state spins, the \SIrange{500}{800}{\mega\hertz} resonance of radical pairs in biological qubits, and any future qubit resonance in the GHz range. Live cell imaging with simultaneous biological and solid state nanoparticle qubits in the same cell demonstrates the power of this technique for multiplexed quantum sensing. We anticipate this approach will open new opportunities for researchers to explore quantum sensing in live cells, and, ultimately, entanglement between a solid-state qubit and a protein-hosted spin qubit.
\end{abstract}
\maketitle

\section{Introduction}
\label{sec:Intro}
Quantum sensors have moved from physics demonstrations to laboratory workhorses over the past decade, and the most consequential advances now sit at the interface between materials-physics qubits and biology. Two parallel lines of progress define the current frontier. On the biological side, spin-correlated radical pairs (SCRPs)~\cite{Hore2016, Steiner1989} and, more recently, genetically encoded fluorescent-protein spin qubits~\cite{Abrahams2026,Feder2025,Burke2026,Ingaramo2024,Burd2026, Xiang2025, Ross2026} have established that a coherent spin degree of freedom can be embedded --- and read out optically --- directly inside a living cell.  On the solid-state side, the negatively charged nitrogen-vacancy (NV) center in diamond has become the archetypal room-temperature optically addressable spin qubit, with optical initialization and readout enabling nanoscale magnetometry, thermometry, and electrometry~\cite{Doherty2013, Schirhagl2014, Degen2017}.

\begin{figure*}[t]
  \centering
  \includegraphics[width=\linewidth]{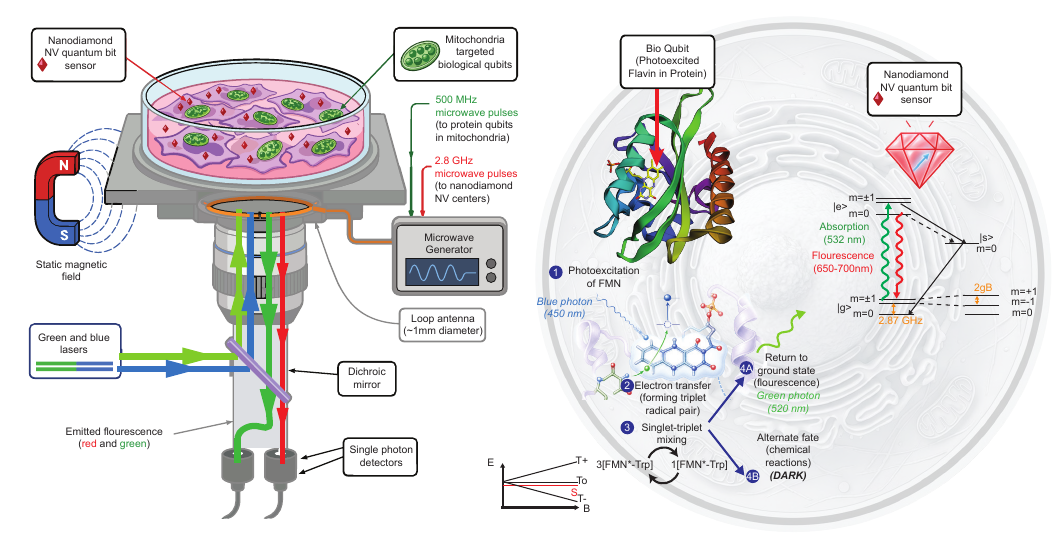}
  \caption{Conceptual diagram of pulsed quantum microscope showing biological and solid state qubits in the same cell in one system. Full quantum state manipulation of both qubits is enabled by pulsed optics, single photon detectors, and pulsed microwave sources (left). Schematic of the energy levels for the two particular quantum bits demonstrated in this work: Biological quantum bit MagLOV and the associated Jablonski energy levels, showing the flavin absorption and emission events, the singlet-triplet mixing, and effect of magnetic field, as well as the well-known crystal structure of an NV defect in diamond, and the associated Jablonski energy levels (right). Generative AI was used to help in preparation of this figure. Figurelabs.ai (GPT Image 2 model) was used to generate an initial draft, with subsequent modifications by the authors in Adobe Illustrator. }
  \label{fig:SchematicWithQuantumBits}
\end{figure*}

Within the solid-state arm, the most important recent development is that entanglement has been converted from a resource for quantum information into a resource for sensing. Superresolution imaging of 2 NVs for gradient magnetometry was demonstrated by one of us in 2010~\cite{Shin2010}, the same year entangled spin pairs were demonstrated by Neumann et al.~\cite{Neumann2010_2}. Recently theory and experiments have been developed to exploit our initial findings for dramatically improved sensitivity. Rovny et al.~\cite{Rovny2025} prepared Bell states in a pair of NV centers separated by $\sim$10 nm and showed that the signal-to-noise ratio for a correlated magnetic-field observable improves by up to $\sim$30$\times$ over a non-interacting two-NV baseline, with a $\sqrt{2}$ quantum-correlation advantage and a tractable decoherence-penalty term. In a complementary experiment, Zhou and colleagues~\cite{Zhou2025} engineered a decoherence-free-subspace state, $(|\Uparrow\Downarrow\rangle + i|\Downarrow\Uparrow\rangle)/\sqrt{2}$, in a \SI{5.2}{\nano\meter} NV pair and reported a 3.4$\times$ sensitivity gain together with a 1.6$\times$ spatial-resolution gain, applied to the detection of individual external electron spins. Le and co-workers\cite{Le2025}, working in the diffraction-limited regime, used covariance magnetometry between two spectrally resolved NV centers within the same optical spot to access noise correlations over the MHz-GHz band at \SI{15}{\nano\tesla\cdot\hertz^{-1/4}}. These three papers, taken together, define the contemporary state of the art for NV-pair quantum sensing. Although the demonstration has required luck since NV pairs are rare birds, advances in materials science and fabrication should enable entangled NV pairs to become a commodity in the next few years. Furthermore, this approach is not limited to NV centers in diamond, as extensive investment in the hundreds of millions of dollars to billions of dollars in spin based solid state qubits in other materials such as silicon is rapidly advancing the ability to entangle spin qubits with arbitrary control~\cite{Dumoulin2026,DeMichielis2023,Hu2025,Gilbert2023,Edlbauer2025}. Although this investment is meant for quantum computing, we anticipate it can be translated to quantum sensing in the future. Thus, the future prospects for entangled sensors seems very promising, and only the tip of the iceberg has been shown so far.

A conspicuous absence from this literature is biology. As in our original work~\cite{Shin2010}, the NV-pair experiments of references~\cite{Rovny2025,Zhou2025,Le2025} are performed on engineered diamond substrates -- pillars, implantation-defined arrays, isotopically purified bulk -- and none has been applied to, or even explicitly proposed for, living biological targets. We have recently taken a first step in this direction by demonstrating a noise-suppressed NV-pair gradiometer protocol designed with biological operation in mind~\cite{Alkahtani2026} where the common-mode rejection of two closely-spaced NVs was exploited to null environmental magnetic noise while preserving sensitivity to local gradients of the kind produced by single biomolecules or cell-surface currents, the operating regime in which any future entangled NV-pair biosensor must work. Additionally, we have demonstrated quantum sensing of magnetic fields with NV centers using an off-the-shelf super-resolution confocal microscope (Zeiss LSM 900 Airyscan) by measuring CW (continuous wave) optically detected magnetic resonance (ODMR) \cite{Noh2025}. The instrument described here is the optical front-end that makes such measurements compatible with inverted, live-cell microscopy in both CW and pulsed regimes.

The biological arm of the field has advanced in parallel. The foundational demonstration is that of Ikeya and Woodward~\cite{Ikeya2021}, who imaged magnetic-field-sensitive endogenous auto-fluorescence in live HeLa cells under sub-\SI{25}{\milli\tesla} static fields, extracted a MARY curve with $B_{1/2} = 18.0 \pm 0.5~\si{\milli\tesla}$ and a saturation magnetic-field effect of \SI{3.7}{\percent}, and localized the signal to mitochondrial flavoprotein SCRPs . That experiment was performed in continuous-wave mode: steady-state photoexcitation combined with a low-frequency modulated static field. Continuous-wave detection has a well-known limitation -- the observed MFE becomes strongly dependent on excitation intensity because the precursor and intermediate populations are forced into a non-universal steady state -- and it offers no time resolution below the tens-of-ms modulation scale. The same group has since solved this problem with two companion techniques, pump--probe (PP) and pump--field--probe (PFP) fluorescence microscopy, in which a pair of nanosecond \SI{450}{\nano\meter} pulses, optionally bracketing a rapidly switched ($<\SI{30}{\nano\second}$) magnetic-field gate, samples the dark-state population and the instantaneous radical-pair lifetime down to $\sim\SI{50}{\nano\second}$~\cite{NoboruIkeya2026}. The Ikeya-Woodward PFP method is, to our knowledge, the only published route to time-resolved radical-pair lifetime / magnetic-field-effect (MFE) spectroscopy at cellular spatial scales, and its conceptual core -- nanosecond-gated optics synchronized to a rapidly switched magnetic field -- has guided the electronic architecture of the instrument reported here.

In parallel, directed evolution has produced genetically encoded proteins whose SCRPs are themselves the sensor. The MagLOV and MagLOV2 family, engineered from the AsLOV2 photoreceptor by C450A mutagenesis followed by iterative selection for large magnetic-field effects, generates a flavin--tryptophan SCRP whose singlet--triplet mixing is coherently driven in continuous-wave RYDMR by $\sim\SI{604}{\mega\hertz}$ radio-frequency fields, yielding RYDMR contrasts of order \SI{50}{\percent} in mammalian cells and bacteria and has recently been extended to in-vivo RYDMR in C. elegans~\cite{Abrahams2026,Antill2026,Burd2026}. Feder and colleagues~\cite{Feder2025}, working with enhanced yellow fluorescent protein, showed that the EYFP photoexcited triplet is a genuine microsecond-coherence spin qubit, with CPMG $T_2 = 16 \pm 2~\si{\micro\second}$, optically activated delayed-fluorescence readout at \SI{912}{\nano\meter}, and coherent microwave control at zero field -- inside live HEK293T cells and \textit{E.~coli} at room temperature. These two classes of biological qubit are complementary: MagLOV provides a tunable, high-contrast RYDMR line in the RF band, whereas EYFP provides a long-lived, microwave-addressable single-spin object. A pulsed microscope capable of handling both classes, and of doing so on the same optical bench as an NV experiment, does not currently exist in the published literature.

Our own group has recently demonstrated continuous-wave MagLOV RYDMR with super-resolution localization of mitochondrially targeted biological qubits in live mammalian cells~\cite{Burke2026}. Those experiments use the same inverted microscope body architecture that underlies the present work, in a commercial off the shelf turnkey confocal microscope, but operate in a CW regime and therefore inherit the steady-state limitations identified in references~\cite{Ikeya2021,NoboruIkeya2026}. The natural next step -- and one of the principal motivations for the instrument reported here -- is the transition from CW to pulsed operation: nanosecond-gated photoexcitation, rapidly switched RF and static fields, and time-resolved detection of the fluorescence response. The platform described in the present paper supplies exactly that capability, and does so in a geometry that is simultaneously compatible with NV-center spin control.

The broader significance of unifying these capabilities is that it opens, for the first time, a credible experimental path to entanglement across the solid-state--biological boundary. Two-qubit entanglement between NV centers has now been demonstrated~\cite{Rovny2025,Zhou2025}; two-qubit entanglement between biological qubits -- whether two MagLOV SCRPs in adjacent proteins or two EYFP triplets in neighbouring chromophores -- has not been demonstrated but is not ruled out by any known physics. The experiment that would matter most, and that no existing platform can address, is the hybrid one: a single pulsed microscope capable of preparing, coherently manipulating, and reading out an entangled state shared between a solid-state NV qubit in a surface-proximal nanodiamond and a biological qubit hosted in a protein in the adjacent cell. Such a hybrid-entangled quantum sensor would merge the long coherence and microwave addressability of the NV center with the genetic targetability and intracellular localization of a fluorescent-protein or MagLOV qubit, and would constitute, in our view, the true holy grail of the quantum interface with biology.

In this paper, we report the design, implementation, calibration, and  validation of a pulsed quantum microscope that targets this regime as illustrated in Figure~\ref{fig:SchematicWithQuantumBits}. Our specific contributions are four. First, we demonstrate a stationary-sample, galvo-galvo, inverted-microscope architecture that supports the green-excitation / red-collection / microwave-driven protocols required for NV-center ODMR and $T_1$ spectroscopy, while preserving an optical and sample geometry that can be extended to blue-excitation, nanosecond-gated, RF-driven biological-qubit measurements. Second, we benchmark NV performance on an implanted diamond plate mounted on the stationary Olympus IX71 stage, demonstrating single NV ODMR near \SI{2.87}{\giga\hertz} and $T_1 \approx \SI{1.3}{\milli\second}$ under ambient conditions, which establishes the NV baseline required for future NV-pair sensing protocols inspired by references~\cite{Rovny2025,Zhou2025,Le2025,Alkahtani2026}. Third, we demonstrate biological fluorescence compatibility by imaging TMRE-stained and MagLOV-expressing HeLa cells after limited changes to the excitation and detection paths. Four, we demonstrate, for the first time, simultaneous live-cell imaging of biological and solid state qubits in the same cell. These results establish the microscope-level infrastructure needed to pursue future time-resolved RYDMR, NV-pair sensing, and solid-state--biological quantum-interface experiments.

\section{System Design}
\label{sec:System Design}
\subsection{Confocal Microscope Optical Setup: Imaging}
\label{sec:optical}
\begin{figure*}[t]
  \centering
    \includegraphics[width=\linewidth]{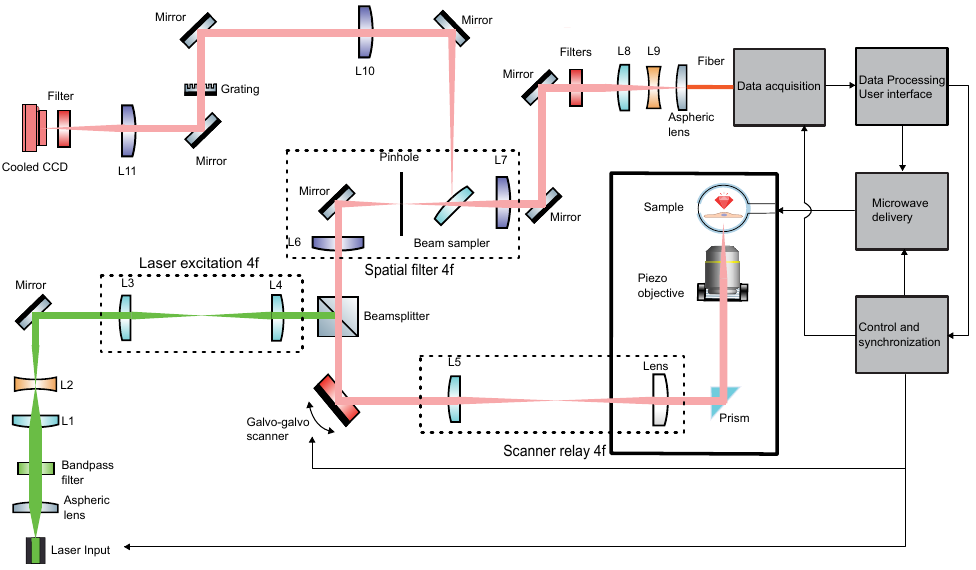}
    \caption{Schematic of the pulsed quantum microscope platform. The optical path comprises a fiber-coupled \SI{520}{\nano\meter} laser, four nested $4f$ relay systems, a galvo-galvo beam scanner, a confocal pinhole, and a single-photon detector (SPD). The microwave delivery routes a nanosecond-gated signal from the RF generator through a power amplifier and PCB loop antenna positioned beneath the sample. Data acquisition is synchronized with both the optical and microwave channels.}
  \label{fig:simplifiedschematic}
\end{figure*}

In order to optimize the confocal microscope, it is important to be able to adjust the laser focus position relative to the microscope focus without using the microscope's own adjustments. A $4-f$ imaging system is well suited to this end. It works by imaging a conjugate plane — where all beam adjustments are made — onto the back aperture of the objective via the galvo mirrors. In this way, changes in angle or focus do not alter the laser beam size or position on the back aperture. The typical adjustments consist of mirror on a tilt plate for laser beam $x$, $y$ angles which map to $x$, $y$ position in the object plane, and a telescope to adjust convergence/divergence of the laser beam which maps to $z$ position. The output lens of the telescope is as close as possible to the mirror, so that both are imaged onto the back aperture. The use of a $4-f$ imaging system ensures that both laser beam position and curvature are imaged correctly.

The basic design of a galvo-galvo based confocal microscope involves four nested conjugate image planes, as shown in Figure~\ref{fig:simplifiedschematic}. The first is defined by the microscope manufacturer through the objective and internal field-lens combination. The second relays this image to a conjugate plane containing the confocal pinhole. The third is a nested imaging system that images the galvo mirrors onto the back aperture of the microscope objective, so that galvo scanning does not cause the laser beam to miss the objective pupil. In some implementations, the galvo may also be used with a matched $f-\theta$ lens, which further constrains the relay design. The fourth images a collimated laser beam, with a size matched to the objective back aperture, onto the galvo mirrors so that small alignment adjustments do not cause beam clipping at the scanner. Each of these imaging subsystems requires $x-y$ (tip-tilt) and $z$ (focus) adjustment. In our implementation, each relay is based on a $4f$ design, which not only forms the required image plane but also preserves the appropriate beam curvature at that plane. Although there is some redundancy between the $x-y-z$ adjustments of the microscope and those of the pinhole, we find that this additional flexibility greatly simplifies both initial alignment and subsequent optimization.

The schematic implementation of this design in our system is shown in Figure~\ref{fig:simplifiedschematic}. The confocal microscope platform is constructed on an Olympus IX71 inverted microscope. This microscope base is the most commonly used in live-cell imaging in biology, and serves as the mechanical platform for dual solid state and biological qubit imaging in this paper. The setup combines a dual color \SI{520}{\nano\meter} and \SI{450}{\nano\meter} laser excitation path, a galvo-galvo based scanning module, and a single-photon detection path through three carefully aligned $4f$ imaging systems. Each $4f$ relay plays a distinct role in maintaining beam conjugation and minimizing aberrations across the excitation, scanning, and detection pathways. A detailed schematic and parts list are provided in Appendix~\ref{appendix:Alignment Procedure} and Appendix~\ref{appendix:parts list}, respectively.

\begin{figure}[h]
  \centering
  \includegraphics[width=1\columnwidth]{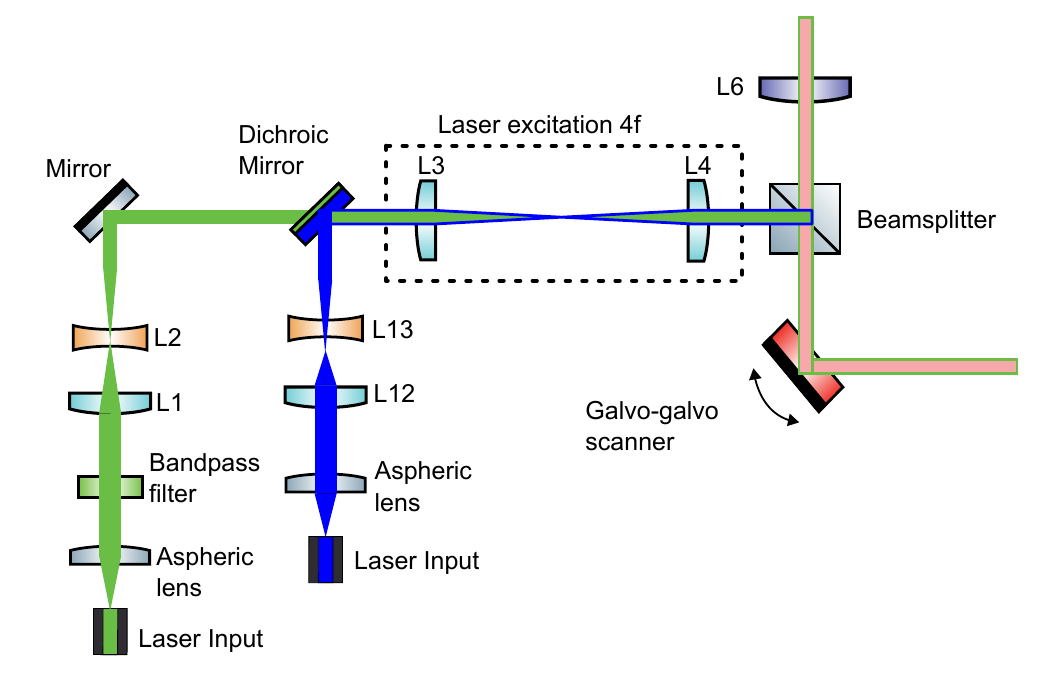}
  \caption{Dual-excitation configuration at the laser input. A dichroic mirror combines the fiber-coupled \SI{520}{\nano\meter} (NV) and \SI{450}{\nano\meter} (MagLOV) excitation  beams into a single collinear path routed through the shared $4f$ relay, galvo-galvo scanner, and microscope objective.}
  \label{fig:dualcolorconfiguration}
\end{figure}

Excitation is provided by a single-mode fiber-coupled \SI{520}{\nano\meter} laser for the solid state qubits, and a \SI{450}{\nano\meter} fiber-coupled laser for the biological qubits. The detailed laser path is shown in Figure~\ref{fig:dualcolorconfiguration}. The excitation light is collimated and guided into the beam-scanning module through the first $4f$ relay. This relay consists of two convex lenses and forms a beam waist at the plane of the galvo-galvo scanner (Thorlabs LSKGG4). The scanner comprises two orthogonally mounted galvanometric mirrors capable of independent angular deflection in the $x$ and $y$ directions. These mirrors introduce time-varying angular shifts into the excitation beam, which are then relayed to scan the beam laterally across the sample.

As shown in Figure~\ref{fig:dualcolorconfiguration}, the two excitation beams are combined at the input of the first $4f$ relay using a dichroic mirror. The existing \SI{520}{\nano\meter} path, which was aligned first for NV confocal imaging, passes through a bandpass filter before combination. The combined beam is then routed through the same optics, galvo-galvo scanner, second $4f$ relay and objective. To convert the angular deflection from the galvo-galvo scanner into lateral displacement at the sample, a second $4f$ relay is used. This relay is composed of two more convex lenses and images the scanned beam onto the back aperture of the microscope objective. In this configuration, the angular deviation imparted by the galvo mirrors is transformed into an approximately linear lateral translation of the beam at the sample plane. Provided that the scanned beam adequately fills the objective aperture, this configuration supports diffraction-limited point excitation across the scan field. The focused beam is delivered to the sample through a 100$\times$ high numerical aperture (NA) oil objective lens (Zeiss Alpha Plan-Apo 100x, NA 1.46). Fine axial focusing is performed by moving the objective with a piezoelectric $z$-stage (Thorlabs PFM450E), which provides closed-loop control over a \SI{450}{\micro\meter} travel range with \SI{3}{\nano\meter} resolution. This piezo-based control is essential for axial sectioning and for optimizing signal collection from nanoscale NV emitters in the solid state or as nanodiamonds in live cells, and confocal live-cell imaging of biological qubits.

The fluorescence is collected through the same objective and descanned by the galvo mirrors, ensuring that the emission path is optically matched to the excitation path in reverse. The beam is then reflected by a beamsplitter, which separates the red-shifted NV emission from the excitation beam. In our implementation, we use an 10:90 beamsplitter to increase fluorescence throughput. Many reported NV microscope designs instead use a dichroic mirror at this stage, but we find that the beamsplitter-based approach significantly simplifies alignment while causing only a small reduction in detected signal. The emission is then focused through a pinhole located at a plane conjugate to the sample, for this purpose we used an achromatic doublet (Thorlabs AC254-150-AB) which is optimized to provide a nearly constant focal length across a broad bandwidth by minimizing the chromatic aberration of the lens. This spatial filter removes out-of-focus background and enforces axial resolution by rejecting light that does not originate from the focal plane.

After the pinhole, the beam is relayed through a third $4f$ system, composed of two lenses, which images the pinhole plane onto the entrance of a multimode optical fiber. This fiber delivers the signal to a single-photon avalanche diode detector (SPD, Excelitas SPCM-AQRH-10-FC). A long-pass emission filter is placed before the fiber to further suppress residual excitation light. The TTL output pulses from the SPD are recorded by a Swabian Time Tagger 20, which performs high-resolution time-stamped photon counting for both raster-scanned confocal imaging and time-resolved fluorescence measurements.

As noted above, the use of three external $4f$ relay systems is critical for maintaining optical conjugation and alignment throughout the setup. The first $4f$ system ensures that the laser beam is properly imaged onto the galvo scanner, preserving both collimation and $x-y$ position to avoid clipping on the small (\SI{4}{\milli\meter}) galvo mirrors. The second $4f$ system projects galvo deflection onto the objective back aperture to achieve precise lateral beam scanning. The third $4f$ system ensures that the fluorescence signal exiting the pinhole is efficiently coupled to the detector with minimal aberration and signal loss. A spectroscopy arm is also incorporated through a related relay stage that includes a diffraction grating mirror.

The scanning field of view is defined by the galvo-galvo scanner scan range, typically operated below $\pm\SI{2}{\volt}$ on each axis. Calibration using a microscope resolution target yields a conversion factor of approximately \SI{24}{\micro\metre\per\volt}, corresponding to a maximum field of view of about \SI{48}{\micro\metre}. With optimized alignment and stable excitation, the system is capable of sub-micron, diffraction limited spatial localization of solid state spin qubit and biological qubit fluorescence. 

\begin{figure}[h]
  \centering
  \includegraphics[width=1\columnwidth]{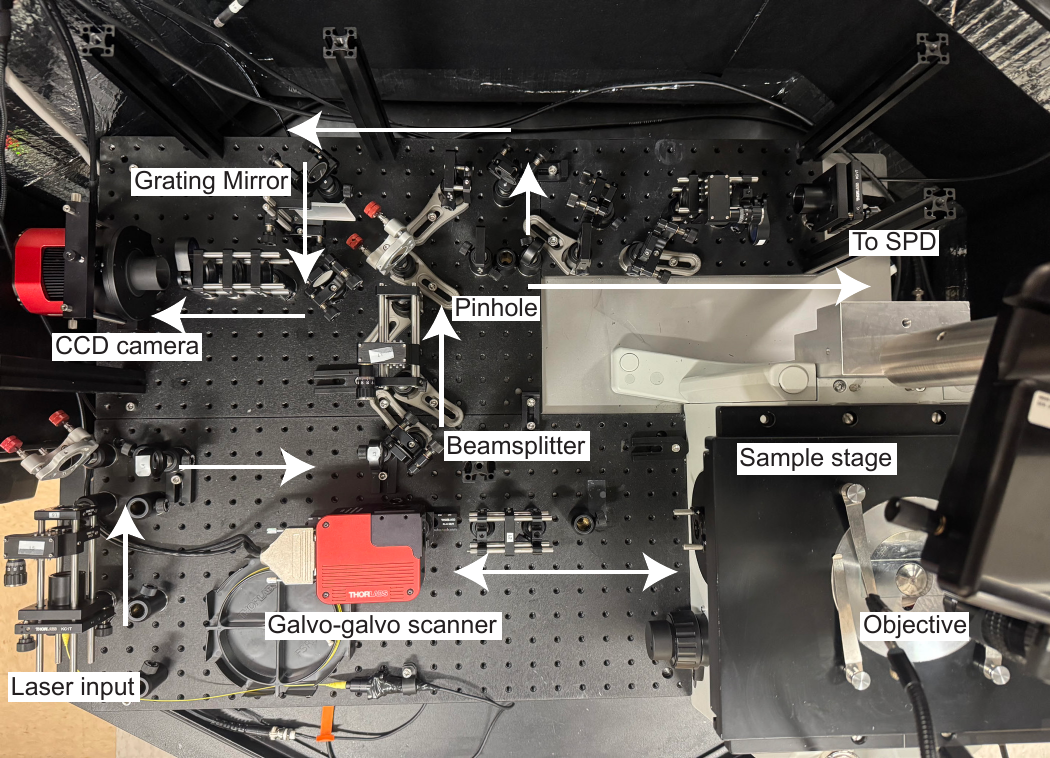}
  \caption{Photograph of the pulsed quantum microscope showing the inverted Olympus IX71 microscope base (right) and assembled optical breadboard (left).}
  \label{fig:olympushardware}
\end{figure}

The system was assembled inside an isolated cabinet and on top of an anti-vibration table to reduce external optical noise and vibrations. Figure~\ref{fig:olympushardware} shows a photograph of the completed bench top instrument, in which the galvo-galvo scanner, laser excitation path, and single-photon confocal detection module are mounted on an Olympus IX71 inverted microscope for both solid state and biological qubit imaging.

\subsection{Confocal Microscope Optical Setup: Optical pulsing}

Pulsed quantum state manipulation requires that excitation is delivered in short, well-defined bursts with high extinction ratio, synchronized to the MW fields and data acquisition. Two approaches were tested with our optical bench: free-space using acousto-optic modulators (AOMs), and electronic modulation of the fiber-coupled laser diode drive current. We implemented and benchmarked both during platform development; the relative merits, alignment complexity, and cost are discussed below.

\subsubsection{AOM based optical modulation}

\begin{figure}[h]
  \centering
  \includegraphics[width=1\columnwidth]{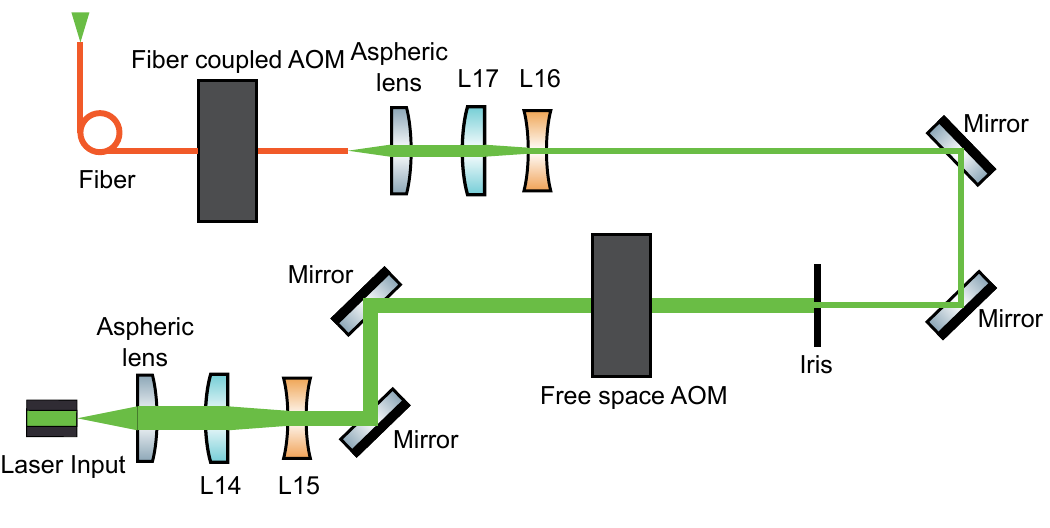}
  \caption{Schematic of the dual AOM-based laser pulse configuration.}
  \label{fig:AOMconfigurationSchematic}
\end{figure}

Most AOMs have an extinction ratio of \SI{40}{\decibel} which,  while sufficient for coherent quantum state manipulation such as Rabi oscillations, are not sufficient for $T_1$ lifetime measurements and, ultimately, entanglement, since the small "off" laser power can contaminate the quantum state distribution in sensitive coherence protocols. Some papers~\cite{Yuan2024} are satisfied with a single AOM as it can enable some but not all quantum state manipulation. For more demanding quantum state manipulation, most researchers use two AOMs in series or add a double pass setup, giving around \SI{80}{\decibel} of extinction ratio, which we have implemented here by combining a free space AOM (Isomet 1205C-2-790) and a fiber coupled AOM (Brimrose TEM-200-25-20-532-2FP), as shown in Figure~\ref{fig:AOMconfigurationSchematic}. However, that gives extra complexity, requiring us an extra external optical board for the free space AOM section, and cost, since commercial AOMs are of order a few thousand dollars. We have found recently that commercial AOMs are globally out of stock or no longer manufactured, with months to years of lead time. Used AOMs have also been demonstrated in this work, but with mixed results, as high optical powers can damage the crystals giving visible burn marks in the crystal. One company promised \SI{60}{\decibel} of extinction ratio in a single fiber coupled AOM, which we purchased for 5000 USD, but the product (even after a 6 months lead time) did not function properly, was not repaired for many months after sending back to the company, and is no longer offered commercially with that design. Finally, AOMs must be aligned properly in the optical system, and fiber coupled AOMs need care in the fiber mounting and fiber-free space conversion and are not broadband compatible which limits their use for multiple color configuration but it is enough for people working only with solid state qubits as NV-centers. 

\subsubsection{Electronic optical modulation}

Most modern lasers function as solid state diodes, whose current can be modulated to modulate the optical power. The advantage is the simplicity without the need for additional optics to align and low cost. The disadvantage is the complexity of the current driver circuit for high speed, high current pulses. In addition, laser stability may be inferior when the drive current is not steady, due to thermal drift and transients. Nothing will be more steady than a laser that is turned on in the morning, warmed up, and allowed to stabilize before the run is started.  However, we have found the electronic pulsed lasers are more than sufficient for the applications described in this paper. We demonstrate and provide to the community as part of this work an open source current driver that can be manufactured for about 20 USD by a commercial PCB foundry in about a week (including manufacturing and global delivery to anywhere in the world). The detailed design is provided in Appendix~\ref{appendix:LaserDiodeCurrentDriver}.

\subsection{Microwave System}

The system is modular so that different RF subsystems can be swapped out. A general feature is the delivery of high power microwaves to the spin system connected to a pulse electronics setup, synchronized under computer control with the laser pulsing and photon detection setup.

\subsubsection{Microwave delivery}

\begin{figure}[htbp]
  \centering
  \begin{subfigure}{\columnwidth}
    \centering
    \includegraphics[width=\columnwidth]{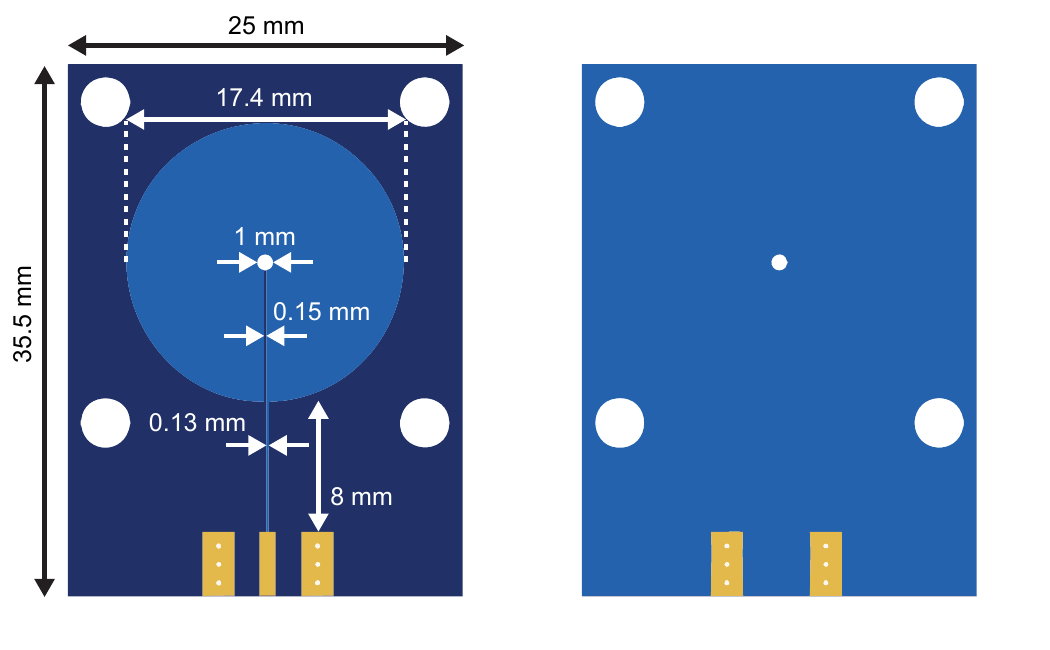}
    \caption{Resonant antenna design.}
    \label{fig:AntennaDesign}
  \end{subfigure}
  \begin{subfigure}{\columnwidth}
    \centering
    \includegraphics[width=\columnwidth]{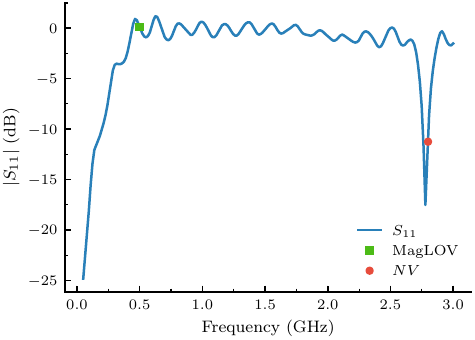}
    \caption{$|S_{11}|$ plot of the resonant antenna design.}
    \label{fig:S11ResonantAntenna}
  \end{subfigure}
  \caption{Antenna design and $|S_{11}|$ plot of the resonant antenna.}
  \label{fig:AntennaDesignAndS11}
\end{figure}

Three different designs were benchmarked for microwave delivery at the sample for experiments done with this system. The first was a resonant antenna, designed specifically for NV center resonance of \SI{2.87}{\giga\hertz}, based on the paper by Misonou~\cite{Misonou2020}. Figure~\ref{fig:AntennaDesign} shows the geometry of the antenna and Figure~\ref{fig:S11ResonantAntenna} the $|S_{11}|$ parameter measured in the laboratory. The advantage of the resonant design is enhanced AC magnetic field at a given power level. The PCB antenna design was adapted from a previously reported layout optimized for quantum control experiments~\cite{Mariani2022}. 

\begin{figure}[htbp]
  \centering
  \begin{subfigure}{\columnwidth}
    \centering
    \includegraphics[width=\columnwidth]{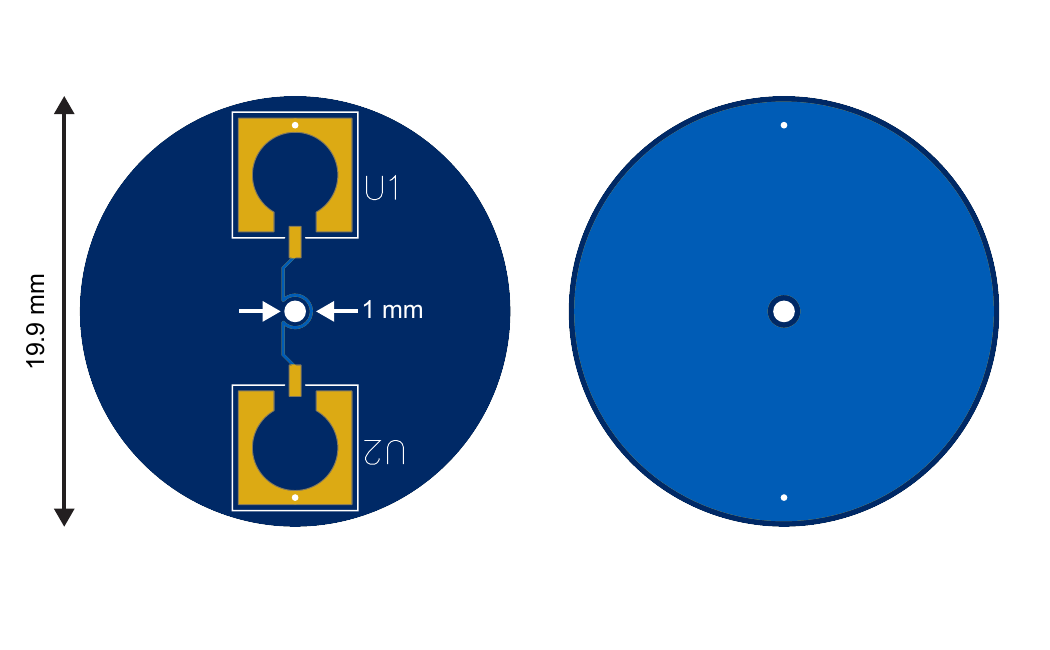}
    \caption{Broadband antenna design for petri dish.}
    \label{fig:AntennaDesignBroadband}
  \end{subfigure}
  \begin{subfigure}{\columnwidth}
    \centering
    \includegraphics[width=\columnwidth]{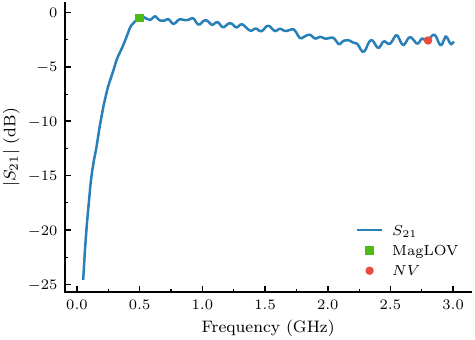}
    \caption{$|S_{21}|$ plot of the broadband antenna for petri dish designed in this work.}
    \label{fig:AntennaS21Broadband}
  \end{subfigure}
  \caption{Antenna design and $|S_{21}|$ plot of the broadband antenna for petri dish designed in this work.}
  \label{fig:AntennaDesignAndS21Broadband}
\end{figure}

The second antenna used in this work was previously reported by Yuan~\cite{Yuan2024} and consist of a broadband \SI{1}{\milli\meter}-loop transmission line. These design work properly for solid state qubits. However, they are not compatible with biological qubits. Therefore, a broadband transmission line compatible with petri dish for cell mounting was designed, which consisted of a \SI{1}{\milli\meter}-loop in a \SI{50}{\ohm} transmission line, the broadband $|S_{21}|$ flat response, as shown in Figure~\ref{fig:AntennaDesignAndS21Broadband}, makes the proposed design compatible with both solid state and biological qubit. This design is fully compatible with petri dishes 35 mm glass-bottom (No. 1.5, MatTek).

\subsubsection{Solid state qubit RF system}
Spin transitions in the NV centers are driven by microwave fields generated and routed through a custom RF chain terminating in a planar loop antenna, as shown in Figure~\ref{fig:simplifiedschematic}~\cite{Mariani2022}. The chain begins with an RF signal generator (RIGOL DSG836), then that signal is fed into an I/Q Modulator (Texas Instruments TRF3705) where the TTL signal from Pulse Streamer enables time-resolved microwave pulse modulation with nanosecond-scale timing control. 

The gated microwave signal is then amplified by a high-power broadband amplifier (Mini-Circuits ZHL-16W-43-S+), which delivers up to \SI{16}{\watt} over the \SIrange{1.8}{4.0}{\giga\hertz} range. To protect the amplifier from reflected power, a microwave circulator (Pasternack PE83CR005) is placed at its output. Then the amplified signal is passed through a step attenuator (Narda 4705B-99), which provides manual control of the microwave amplitude. The attenuator's output is then routed to a printed circuit board (PCB) loop antenna positioned directly beneath the sample.

The antenna is designed to generate an oscillating microwave magnetic field ($B_1$) that couples efficiently to the NV spin transition. The diamond sample is mounted immediately above the center of the loop, in proximity to the antenna surface while remaining optically accessible from above and electrically isolated from the microwave structure. This geometry provides strong local microwave coupling while preserving the optical working distance required for confocal excitation and fluorescence collection.

\subsubsection{Biological qubit RF system}

The biological qubit system also can undergo ODMR, as we have recently demonstrated in reference \cite{Burke2026}. Since the ODMR frequency for MagLOV cells is around \SI{500}{\mega\hertz}, the same setup is used, but the microwave amplifier needs to be replaced by one with the adequate bandwidth (Mini-Circuits ZHL-10M2G0005+) which in the addition to the petri dish compatible loop transmission line designed here allows us to drive the spin transitions of biological qubits. 

\subsubsection{Imaging and qubit quantum control}

All the above use a CW generator, then a post-generator microwave I/Q modulator turns the RF on and off. The pulses fed to the switch are generated by the Swabian Pulse Streamer 8/2, which is also used to gate the photon detection and laser. An alternative is to use an FPGA based DAC, and synthesize the RF frequency directly in the digital domain. We have also implemented this solution for the NV centers, using an RFSoC 4x2 board from RealDigital/AMD~\cite{AMD_RFSoC4x2} which integrates a Zynq Ultrascale+ RFSoC XCZU48DR-2FFVG1517E FPGA. The 2 integrated DACs can synthesize up to \SI{9}{\giga\byte\per\second} at 14 bits, so can create up to \SI{5}{\giga\hertz} microwave signals under arbitrary control. An additional advantage is that the photon pulse counting is also completely handled by the RFSoC4x2 board. Finally, the cost is much lower, around 2000 USD for the RFSoC4x2 board, compared to around 18000 USD for Time Tagger, pulse streamer and signal generator solution.

Spin manipulation has been already demonstrated for NV centers elsewhere using QICK-DAWG~\cite{Riendeau2023} (a fork of QICK\cite{Stefanazzi2022}). Additionally, we demonstrate here for the first time its use for confocal imaging as shown in Appendix~\ref{appendix:RFSoC4x2}. As our antenna is broadband, using the FPGA solution enables arbitrary RF delivery to both quantum bits in live cells.

\subsection{Solid state qubits}

NV centers in diamond emit photoluminescence (PL) under green laser excitation. This emission includes a sharp zero-phonon line (ZPL) at \SI{637}{\nano\meter} and a broad phonon sideband extending to approximately \SI{800}{\nano\meter}~\cite{Alkauskas2014}. The ZPL corresponds to the direct optical transition between the ground and excited electronic states of the negatively charged NV center and serves as a key spectral signature for identifying NV PL~\cite{Sewani2020}.

Several types of diamond samples were used to evaluate the platform across different PL and spin-measurement conditions. For nanodiamond-based measurements, dispersions containing \SI{70}{\nano\meter} nanodiamonds (Adamas, NDNV70nmHi10mL, \(\sim\)100 NVs per particle), \SI{40}{\nano\meter} nanodiamonds (Adamas, NDNV40nmHi1mL, \(\sim\)10 NVs per particle) and \SI{10}{\nano\meter} nanodiamonds (Adamas, NDNV10nmMd1mL, \(\sim\)1 NV per particle) were prepared for confocal imaging and low-density emitter measurements. Equal volumes of the two dispersions were mixed in a 1:19 ratio and then further diluted in a 1~wt\% aqueous solution of polyvinyl alcohol (PVA) to improve dispersion uniformity and adhesion to the substrate. Prior to deposition, the mixture was vortexed and bath-sonicated for 10~min to promote homogeneous suspension. The resulting solution was deposited onto a quartz microscope slide and spin-coated at 3000~rpm for \SI{20}{\second}, followed by baking on a hot plate at \SI{80}{\degreeCelsius} for \SI{10}{\minute}. This procedure produced a well-separated layer of nanodiamonds suitable for confocal imaging and optical targeting of isolated emitters. 

For ensemble NV measurements, \SI{15}{\micro\meter} and \SI{150}{\micro\meter} diamond powders (Adamas, MDNV15umHi30mg and MDNV150umHi30mg, NV concentration 3~ppm) were used. A thin layer of cyanoacrylate adhesive (Starbond) was applied to a glass coverslip, and a small amount of NV powder was drop-cast onto the surface. After drying, the coverslip was attached to the PCB antenna, providing stable sample mounting while maintaining compatibility with both optical excitation and microwave delivery. These nanodiamond and powder samples were used during initial platform alignment, confocal scanning calibration, PL-throughput optimization, and validation of the system's ability to locate and image NV emitters across a range of particle sizes. They were not used for the quantitative NV measurements reported below. 

For single-NV measurements on a bulk diamond substrate, we used an implanted diamond plate from Institute for Quantum Optics - Universit\"at Ulm. According to the implantation map for this sample, the substrate is an IIa electronic-grade single-crystal diamond (ELSC) implanted with $^{15}\mathrm{N}^{+}$ ions at an implantation energy of \SI{5}{\kilo\electronvolt} and subsequently annealed in ultra-high vacuum at \SI{1000}{\degreeCelsius} for \SI{3}{\hour}. The implanted side contains multiple circular implantation regions with a nominal diameter of \SI{200}{\micro\meter} and implantation doses ranging from \numrange{1e9}{1e14}\,\si{\per\square\centi\metre}. Low-dose regions, particularly the \SI{1e9}{\per\square\centi\metre} region, were used for isolating single emitters, while higher-dose regions provided brighter fluorescence for survey scans and alignment. Representative characterization conditions for this implanted sample included \SI{300}{\micro\watt} excitation power in front of the objective, and a \SI{650}{\nano\meter} long-pass filter for NV detection. No ND filter was required for the lower-dose regions, whereas ND filtering was necessary for the highest-dose regions because of the large photon count rates. For the confocal single-emitter localization, ODMR, and $T_1$ relaxation measurements presented in Section~\ref{sec:experiments}, we used the implanted bulk diamond plate described before, which provides isolated emitters with well-defined implantation conditions suitable for controlled spin characterization.

\subsection{Biological qubits}

\begin{figure}[h]
  \centering
    \includegraphics[width=\columnwidth]{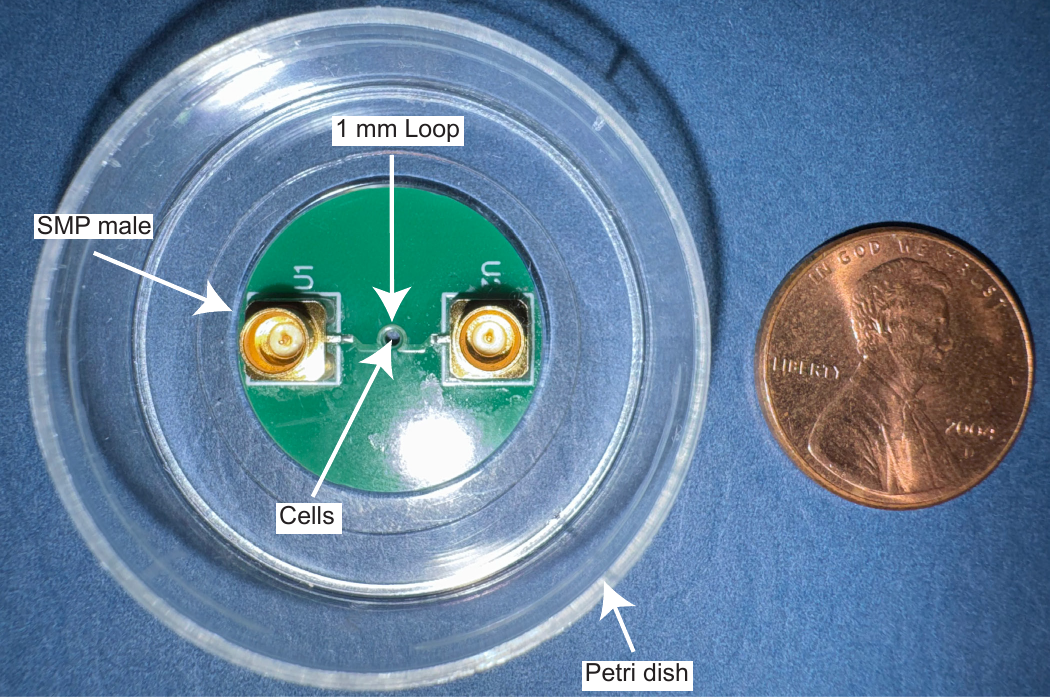}
  \caption{Broadband loop transmission line mounted on a Petri dish. The antenna fits within the coverslip area of the Petri dish, where the cells are cultured.}
  \label{fig:AntennaPetri}
\end{figure}

Directed evolution of the wild type AsLOV proteins has yielded the MagLOV2 family of proteins, whose fluorescence displays strong dependence on applied magnetic fields~\cite{Antill2026,Burd2026, Ingaramo2024, Abrahams2026, Flores2024, Kim2024}. The effect is attributed to a SCRP formed after optical excitation, as sketched for MagLOV in Figure~\ref{fig:SchematicWithQuantumBits}. Absorption of blue light promotes the flavin chromophore, after which photoinduced electron transfer generates a donor-acceptor radical pair. The pair may either separate or recombine to repopulate the emissive singlet state and produce fluorescence. Nuclear hyperfine interactions split the entangled radical-pair spin states into singlet and triplet manifolds that evolve at different rates in a static magnetic field, producing a magnetic-field-sensitive fluorescence yield. In MagLOV2, the non-covalently bound flavin mononucleotide (FMN) cofactor participates directly in this photocycle, with protein-side-chain donors within the AsLOV scaffold serving as the electron donor and FMN as the acceptor and primary emitter~\cite{Burd2026,Antill2026}.

For live-cell experiments, we designed a mitochondria-localized derivative, mtMagLOV2, derived from the R10 Oxford MagLOV2 variant~\cite{Abrahams2026} after ten rounds of directed evolution as described in~\cite{Burke2026}.

\begin{figure}[h]
  \centering
    \includegraphics[width=\columnwidth]{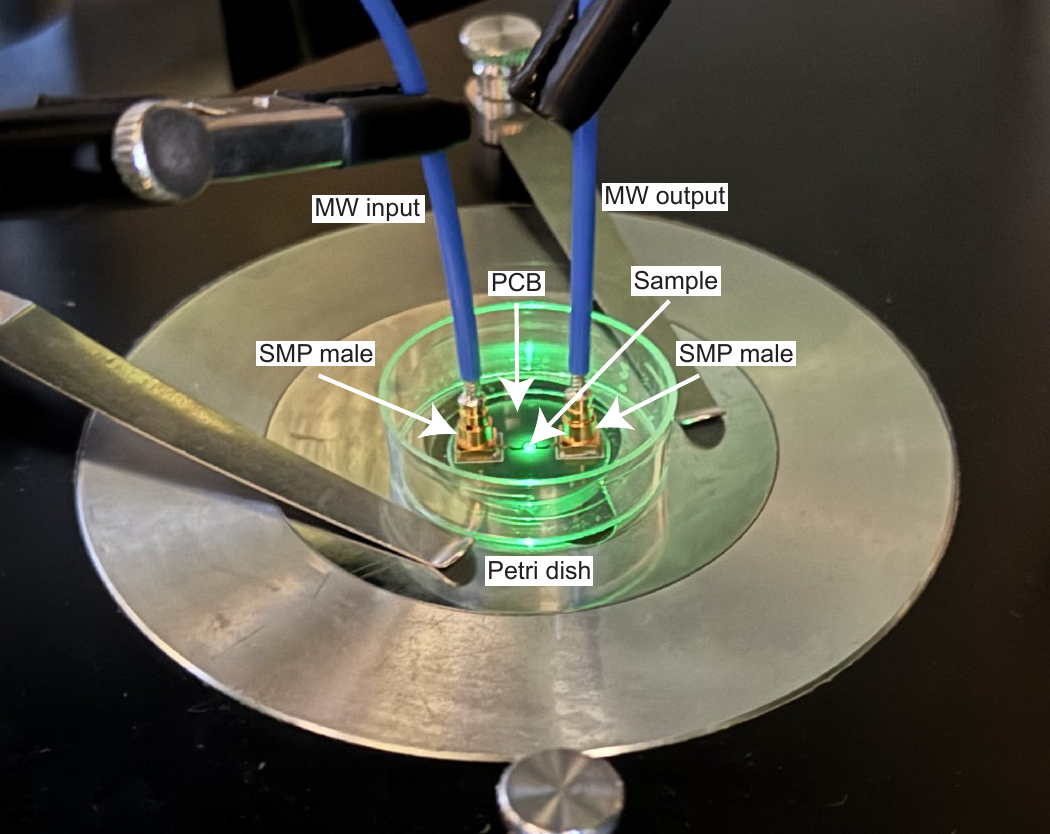}
  \caption{Schematic diagram of the sample mounting for experiments with biological qubits using the designed loop transmission line compatible with petri dishes.}
  \label{fig:BioSampleSchematic}
\end{figure}

The fabricated loop transmission line prototype is shown in Figure~\ref{fig:AntennaPetri}, illustrating how the PCB is integrated with a standard Petri dish while fitting entirely within the coverslip area used for cell culture. Figure~\ref{fig:BioSampleSchematic} shows the sample mounting for experiments with biological qubits using the designed loop transmission line compatible with petri dishes. 

\subsection{Software}

The confocal scanning system is operated through a custom Python-based software suite designed to integrate the key hardware components of the platform, including the galvo-galvo scanner, DAQ (NI-USB 6453), single-photon detector (SPD), Time Tagger 20 for photon counting, Pulse Streamer 8/2 for spin control, real-time navigation camera, and piezo objective scanner. The software is built around a graphical user interface (GUI) implemented in the Napari visualization framework and supports real-time imaging, photon-count monitoring, and multiple scan modes. The code is organized in a modular structure so that different acquisition and imaging tasks can be executed within a single interface. The software is available through a public repository at \url{github.com/burkelabuci/Single_NV_Scanning_Microscopy}.

The GUI provides full-field scanning with live feedback as well as zoomed scans of user-selected regions of interest. Scan parameters such as range, resolution, dwell time, and settle time are set directly through the interface. For each acquisition, the software automatically saves the selected scan parameters together with the measured data in timestamped directories, improving reproducibility and simplifying later analysis.

To facilitate alignment and optimization, the software continuously displays real-time photon counts from the SPD in the main GUI. This feature assists in coarse sample positioning, optical alignment, and focal adjustment. For axial optimization, the software controls the piezo objective scanner (Thorlabs PFM450E) to perform automated $z$-sweeps and determine the focal plane with the highest photon count. The $z$-sweep range and step size are user-defined, with a typical operating range of \SIrange{0}{450}{\micro\meter} in \SI{10}{\micro\meter} steps. In addition to two-dimensional raster scans, the software also supports single-axis scans along either $x$ or $y$, which are useful for beam alignment and for generating position-dependent PL profiles during optimization.

The software further supports live and single-frame image acquisition from a camera. Camera operation is handled in a separate thread so that image streaming does not interrupt the confocal acquisition workflow. In practice, the cameras are used primarily for sample navigation, coarse positioning, and emitter pre-localization before photon-count-based scans are performed.

For each confocal scan, the software generates a raster grid based on the user-defined scan parameters and converts this grid into the corresponding voltage sequence for the galvo-galvo scanner controller through the DAQ. For small scan areas, the use of a high-resolution DAQ is advantageous. The sample is then scanned in a row-by-row raster pattern. Because the galvo mirrors must reposition at the beginning of each new scan line, an additional settle time is introduced at the start of each row to suppress image artifacts associated with rapid beam motion. At each scan point, the software records the PL signal using the Time Tagger for a defined bin-width. In the present implementation, the resulting counts are normalized to counts per second for display and analysis.

To improve scan fidelity and acquisition speed, photon counting was synchronized to the raster scan through a hardware-based strategy using the Time Tagger \texttt{CountBetweenMarkers} function together with pre-generated $x$ and $y$ axis scan waveforms loaded into the DAQ buffer. In this configuration, the DAQ clock was routed to the Time Tagger to provide hardware-level synchronization between galvo position and photon-count acquisition. This approach substantially reduced timing jitter, eliminated imaging artifacts observed in earlier software-synchronized implementation, and reduced the minimum time per pixel to $\approx\SI{100}{\micro\second}$, limited primarily by the stepping speed of the galvo-galvo scanner.

The software interface also includes galvo positioning, allowing the user to return to any selected point within the scan field for real-time optimization or detailed follow-up measurements. This functionality is particularly useful when identifying isolated bright spots and revisiting them after higher-resolution scans. An autoscale function is implemented to facilitate scans over different spatial ranges. This feature relies on prior calibration of the galvo scan voltage to the real-space sample coordinates. A calibration microscope slide was used for this purpose. This value is incorporated into the software to enable automatic pixel-to-micron conversion during imaging and analysis.

\section{EXPERIMENTAL DEMONSTRATIONS}
\label{sec:experiments}
\subsection{Confocal Microscopy}

High-resolution confocal imaging was performed using the proposed confocal microscope design. The excitation laser was scanned across the sample by the galvo mirrors, while the resulting PL was collected through the high NA oil-immersion objective, see Section~\ref{sec:optical}. The emitted signal was spatially filtered by a \SI{100}{\micro\metre} pinhole and detected with a single-photon avalanche diode, enabling background rejection and high-sensitivity photon-count-based imaging.

\begin{figure}[h]
  \centering
  \includegraphics[width=\columnwidth]{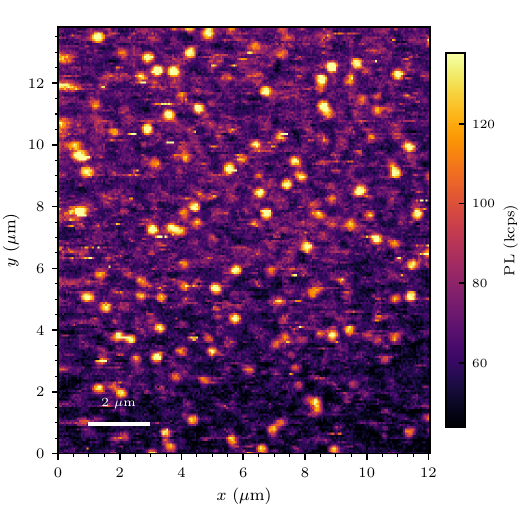}
  \caption{(a) Confocal PL image of a low-dose implanted region (1e9 $^{15}\mathrm{N}^{+}\mathrm{cm}^{-2}$) in the diamond plate. Hot pixels were corrected by median filtering (3×3 kernel).}
  \label{fig:SingleNVScan}
\end{figure}

Figure~\ref{fig:SingleNVScan} shows a representative confocal single NV scan acquired from a low-dose implanted region of the bulk diamond plate. This bulk diamond sample was used to localize isolated emitters for follow-up optical and spin measurements. The confocal image reveals multiple spatially separated bright spots within an approximately \SI{100}{\micro\meter\squared} field of view. 

\begin{figure}[h]
  \centering
  \includegraphics[width=\columnwidth]{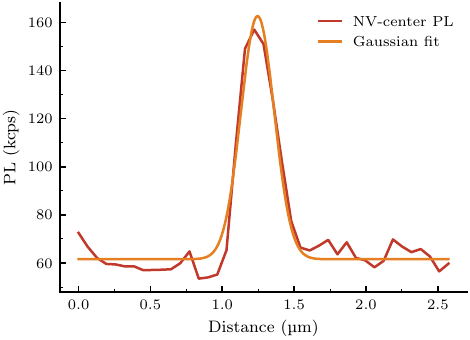}
  \caption{Lateral resolution of the confocal microscope determined from PL intensity profiles of sub-diffraction Single NV spots from Figure~\ref{fig:SingleNVScan}.}
  \label{fig:SingleNVResolution}
\end{figure}

The lateral resolution of the confocal microscope was determined from PL intensity profiles of isolated single NV centers, which behave as point emitters much smaller than the diffraction limit, from Figure~\ref{fig:SingleNVScan}. Line profiles were extracted perpendicular to the scan axis using Fiji/ImageJ~\cite{Schindelin2012}, and the FWHM was evaluated directly from the profile data as shown in Figure~\ref{fig:SingleNVResolution} to extract the resolution. 

The half-maximum intensity was calculated as:

\begin{equation}
    I_{1/2} = I_\mathrm{bg} + \frac{I_\mathrm{max} - I_\mathrm{bg}}{2},
    \label{eq:halfmax}
\end{equation}

\noindent with $I_\mathrm{max} = 1.6\times10^{5}$~counts and $I_\mathrm{bg} = 6.0\times10^{4}$~counts, giving $I_{1/2} = 1.1\times10^{5}$~counts. The lateral FWHM was measured to be approximately \SI{300}{\nano\meter}, in good agreement with the Abbe diffraction limit for \SI{520}{\nano\meter} excitation.

\begin{equation}
    d = \frac{\lambda}{2\,\mathrm{NA}},
    \label{eq:abbe}
\end{equation}

\begin{figure}[h]
  \centering
  \includegraphics[width=\columnwidth]{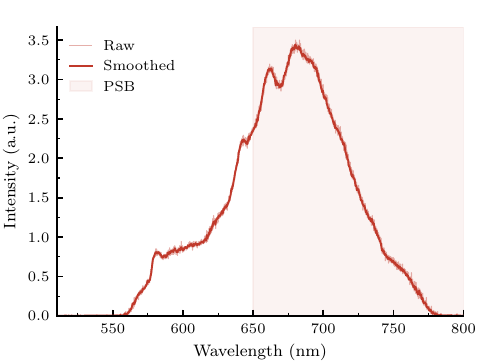}
  \caption{Single NV emission spectrum.}
  \label{fig:SingleNVSpectrum}
\end{figure}

The emission spectrum measured at the marked position is shown in Figure~\ref{fig:SingleNVSpectrum}. The observed spectrum is consistent with NV-related PL, with the NV$^{-}$ and NV$^{0}$ zero-phonon-line reference positions indicated for comparison. It supports identification of the marked spot as an isolated NV center suitable for subsequent ODMR and $T_1$ measurements.

To obtain these data, the sample was first coarsely positioned using the mechanical stage and navigation camera, followed by galvo-based survey and zoom scans to identify regions of interest. Fine optimization was then performed through piezo-assisted $z$ focusing and real-time photon-count monitoring in the software interface. This workflow enabled efficient localization and verification of candidate emitters in the implanted diamond plate and provided the spatial starting point for the spin-resonance measurements presented below.

These results demonstrate that the platform can resolve isolated emitters in a bulk implanted diamond sample. In the present system, confocal imaging serves both as a characterization tool and as an essential targeting step for spectroscopy and spin-resonance experiments on selected emitters.

\subsection{Optically Detected Magnetic Resonance}

\begin{figure}[h]
  \centering
  \includegraphics[width=\columnwidth]{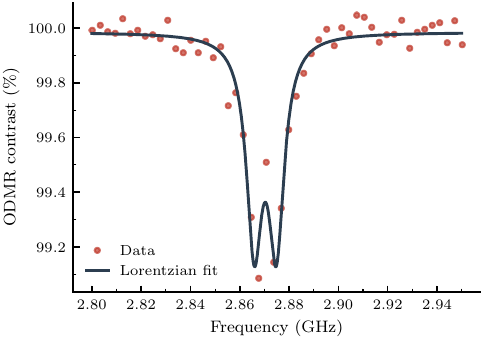}
  \caption{Single NV CW ODMR spectrum. A clear resonance dip is observed near \SI{2.87}{\giga\hertz}.}
  \label{fig:ODMR}
\end{figure}

Continuous-wave optically detected magnetic resonance (CW ODMR) measurements were performed on the selected NV center. During the measurement, the emitter was continuously illuminated with the \SI{520}{\nano\meter} laser while a continuous microwave drive was applied through the PCB antenna. The microwave frequency was swept across the resonance range, and the PL signal was recorded by the single-photon detector. To reduce the effect of slow laser-power drift, the ODMR spectrum was analyzed using the normalized signal-to-reference ratio.

Figure~\ref{fig:ODMR} shows a representative CW ODMR spectrum acquired from the selected single NV center in the implanted diamond sample. A clear PL dip is observed near \SI{2.87}{\giga\hertz}, consistent with the ground-state zero-field spin resonance of an NV center. The measured ODMR contrast was approximately 1\%, and reproducible spectra were obtained across repeated measurements. The limited contrast is attributed to insufficient microwave power delivered to the sample, a consequence of the current mechanical mounting geometry. An improved sample stage is under development to enhance microwave delivery and ensure full compatibility with bulk crystals, nanoparticles, and biological specimens.

This result demonstrates that the platform supports microwave-driven spin readout following confocal localization. In the present workflow, CW ODMR served as the first spin-resonance measurement on the selected center and provided the basis for the subsequent $T_1$ measurement.

\subsection{Spin Relaxation}

\begin{figure}[h]
  \centering
  \begin{subfigure}[b]{\columnwidth}
    \centering
    \includegraphics[width=\textwidth]{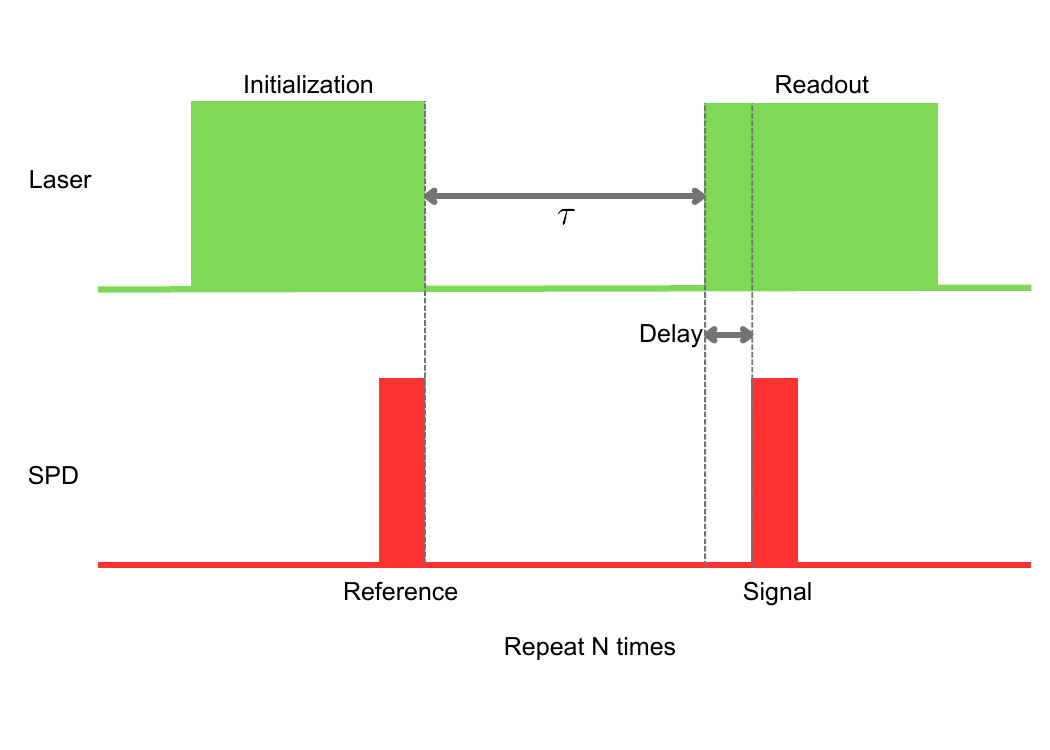}
    \caption{Pulse sequence used for $T_1$ measurement. The first laser pulse initializes the NV spin state and provides the reference window, while the second pulse is used for readout after a variable delay $\tau$.}
  \end{subfigure}
  \begin{subfigure}[b]{\columnwidth}
    \centering
    \includegraphics[width=\textwidth]{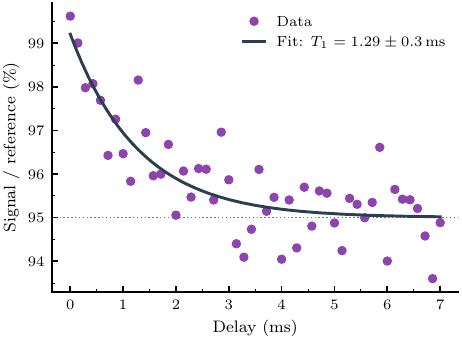}
    \caption{Measured $T_1$ relaxation curve plotted as signal/reference versus delay time, together with a single-exponential fit yielding $T_1 \approx \SI{1.3}{\milli\second}$.}
  \end{subfigure}
  \caption{Spin relaxation measurements on the selected NV center in the implanted diamond sample.}
  \label{fig:T1}
\end{figure}

Spin relaxation measurements were performed on the NV center selected from the confocal scan. The pulse sequence used for this measurement is shown in Figure~\ref{fig:T1}(a). A green laser pulse was first applied to initialize the NV spin state, and photon counts collected near the end of this pulse were used as the reference signal. After a variable dark interval $\tau$, a second laser pulse was applied for readout. The photon-counting window for the readout signal was opened after a delay of \SI{1.5}{\micro\second}, and the ratio of the readout signal to the reference signal was used for normalization in order to reduce sensitivity to slow laser-power fluctuations.

Figure~\ref{fig:T1}(b) shows the measured relaxation curve obtained by varying the delay time $\tau$. As $\tau$ increases, the normalized PL approaches its equilibrium value, consistent with longitudinal spin relaxation under ambient conditions. The data were fitted with an exponential decay function, yielding a relaxation time of $T_1 \approx \SI{1.3}{\milli\second}$.

This result demonstrates that the platform supports time-resolved spin measurements in addition to confocal imaging and ODMR. In the present workflow, the $T_1$ measurements were performed after confocal localization and emitter selection, showing that the microscope platform can be used not only to identify NV centers but also to carry out pulsed spin-relaxation measurements on a selected center.

\subsection{Live cell imaging}

The inverted, stationary-sample confocal architecture was designed to support both solid-state quantum measurements and biological fluorescence samples. To validate this biological imaging channel, we performed fluorescence imaging of HeLa cells. In this configuration, the sample remains fixed on the microscope stage while the excitation beam is raster-scanned by the galvo-galvo scanner.

\begin{figure}[t]
  \centering
  \includegraphics[width=\columnwidth]{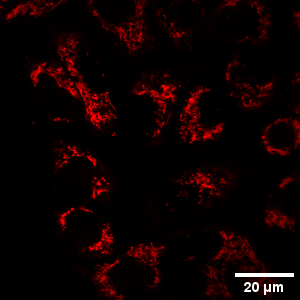}
  \caption{Confocal image of TMRE-stained HeLa cells acquired using the Olympus IX71-based confocal platform.}
  \label{fig:TMRE}
\end{figure}

Representative confocal images of TMRE-stained HeLa cells are shown in Figure~\ref{fig:TMRE}. TMRE is a voltage dye, only live cells will sustain a mitochondrial membrane voltage. The images resolve the cellular morphology and fluorescence distribution across the sample, demonstrating that the platform can accommodate cell-based fluorescence imaging in addition to nanodiamond and bulk-diamond measurements. The inverted microscope geometry is particularly useful for biological samples because it is compatible with coverslips, culture dishes, and microfluidic sample formats.

\subsection{Biological qubit interrogation}

We next applied the platform to HeLa cells expressing MagLOV, a genetically encoded magneto-sensitive fluorescent protein relevant to biological quantum sensing. For this measurement, the excitation source was a Thorlabs LP450-SF25 laser, and the detection filter on the SPD side was replaced with a Thorlabs FBH520-40 bandpass filter to match the MagLOV fluorescence band. The remaining imaging workflow followed the same procedure used for single-NV confocal imaging.

\begin{figure}[t]
  \centering
  \includegraphics[width=\columnwidth]{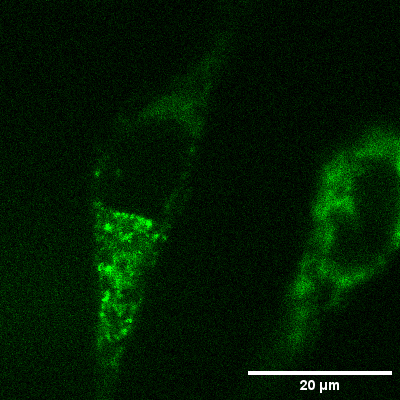}
  \caption{Confocal image of MagLOV-expressing HeLa cells acquired using the proposed Olympus IX71-based confocal platform. }
  \label{fig:MagLOV}
\end{figure}

Figure~\ref{fig:MagLOV} shows a confocal image of the MagLOV-HeLa cells using the proposed Olympus IX71-based confocal platform. This result demonstrates that the same confocal imaging workflow used for NV measurements can resolve MagLOV fluorescence patterns at the cellular scale; the relevant figure of merit for the present platform is single-photon sensitivity and galvo-scan compatibility with the NV-center spin-control workflow, both of which are confirmed here. This result demonstrates that the optical framework developed for NV-center quantum experiments can be reconfigured for genetically encoded biological spin systems through only limited changes to the excitation and detection paths.

Together, these cell-imaging results establish the optical compatibility of the platform with biological fluorescence samples and support its extension toward MagLOV-based ODMR/RYDMR measurements, biological spin-dynamics studies, and future solid-state--biological quantum sensing interfaces.

\subsection{Biological and solid state Qubit Interrogation}

\begin{figure}[h]
  \centering
  \includegraphics[width=\columnwidth]{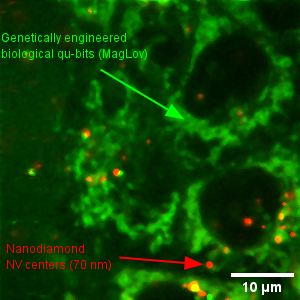}
  \caption{Confocal image of MagLOV HeLa cells with NV solution added to the culture medium.}
  \label{fig:dualcolor}
\end{figure}

In order to demonstrate simultaneous integration of live cells and solid state quantum sensors, we cultured MagLOV HeLa cells in the presence of nanodiamonds. The nanodiamonds were taken up by endocytosis. In Figure~\ref{fig:dualcolor} we show a simultaneous image of nanodiamonds and MagLOV HeLa cells. This demonstrates the first example of live-cell simultaneous solid-state and biolgical qubits in the same system.

\section{Conclusion}
\label{sec:Conclusion}

We have designed, implemented, and initially validated a pulsed quantum microscope built around a commercial Olympus IX71 inverted body that combines solid-state NV-center spin measurements with an optical channel compatible with genetically encoded biological spin systems within a single stationary-sample, galvo-galvo architecture. On the solid-state side, confocal localization of single NV centers in implanted diamond, combined with CW ODMR at \SI{2.87}{\giga\hertz} and a pulsed spin-relaxation measurement yielding $T_1 \approx \SI{1.3}{\milli\second}$, this confirms that the platform operates within the sensitivity and coherence envelope established by recent NV-pair quantum sensing experiments~\cite{Rovny2025, Zhou2025, Le2025}. On the biological side, confocal imaging of TMRE-stained HeLa cells and MagLOV HeLa cells -- a genetically encoded magneto-sensitive fluorescent protein -- demonstrates that the same optical framework supports spatially resolved readout of biological spin systems through only limited reconfiguration of the excitation and detection paths. The coexistence of these two channels on one bench is the central experimental result of this work.

The architectural significance of this coexistence goes beyond convenience. The stationary-sample geometry, in which lateral beam scanning is performed entirely by the galvo-galvo scanner while the sample remains mechanically fixed, is not merely a practical alternative to sample-scanning implementations: it is a prerequisite for the class of experiments that motivates this instrument. Entangled NV-pair sensing~\cite{Rovny2025, Zhou2025}, noise-suppressed gradiometry in biologically relevant field environments~\cite{Alkahtani2026}, and time-resolved RYDMR at cellular spatial scales~\cite{NoboruIkeya2026} all require that microelectrodes, RF coils, magnetic-field delivery structures, and microfluidic or cell-culture interfaces remain undisturbed during acquisition. The present platform satisfies this constraint while simultaneously providing nanosecond-gated optical excitation at \SI{450}{\nano\meter} and \SI{520}{\nano\meter}, rapidly switched static and radio-frequency magnetic fields, single-photon timing with picosecond resolution, and microwave control.

The natural trajectory of this work follows three converging lines. The first is the extension of the NV channel from single-spin to two-spin operation: the Bell-state and decoherence-free-subspace protocols of references~\cite{Rovny2025, Zhou2025} require no hardware additions beyond a second implantation-defined NV site within the confocal volume, and the $T_1$ benchmark reported here confirms that the platform's coherence budget is compatible with those protocols. The second is the transition of the biological channel from continuous-wave to pulsed operation: replacing steady-state photoexcitation with the pump--probe and pump--field--probe nanosecond pulse sequences introduced by Ikeya and Woodward~\cite{NoboruIkeya2026} will unlock time-resolved RYDMR on MagLOV~\cite{Antill2026, Burd2026} and EYFP~\cite{Feder2025} targets with sub-\SI{100}{\nano\second} temporal resolution, removing the steady-state limitation that constrains all current CW biological-qubit experiments including our own~\cite{Burke2026}. The third and most consequential, line is the hybrid regime: a single acquisition in which a solid-state NV qubit in a surface-proximal nanodiamond and a genetically encoded biological qubit in an adjacent cell share a common optical volume and are addressed by the same pulsed sequence. Demonstrating entanglement across this solid-state-to-biological boundary -- merging the long coherence and microwave addressability of the NV center with the genetic targetability and intracellular localisation of a fluorescent-protein qubit -- remains, in our view, the true holy grail of quantum bioimaging, and the instrument reported here is architecturally positioned to pursue it.

\begin{acknowledgments}
This work was supported by the Air Force Office of Scientific Research (AFOSR) under awards FA9550-23-1-0061 and FA9550-23-1-0436, and also in part by the National Science Foundation (NSF) under Award 2153425, and the Noyce Foundation. Any opinions, findings, and conclusions or recommendations expressed in this material are those of the author(s) and do not necessarily reflect the views of the United States Air Force. We thank Prof. Dr. Fedor Jelezko and Jens Fuhrmann from Institute for Quantum Optics - Universit\"at Ulm for providing the ion-implanted NV center samples.

\end{acknowledgments}

\section*{Data Availability Statement}

The data that support the findings of this study are available from the corresponding author upon reasonable request.

\section*{References}

\bibliography{referenceslist}

\clearpage
\appendix
\renewcommand{\thefigure}{\thesection\arabic{figure}}
\renewcommand{\thetable}{\thesection\arabic{table}}

\section{Alignment Procedure}
\label{appendix:Alignment Procedure}
\setcounter{figure}{0}
\setcounter{table}{0}

The alignment procedure consists of three sequential stages: (1) alignment of the microscope, galvo-galvo scanner, and pinhole; (2) alignment of the excitation laser and collection path to the pinhole; and (3) alignment of the pinhole output to the single-photon detector (SPD).

\subsection{Microscope to pinhole alignment}

The purpose of this step is to align the output of the camera port of the Olympus IX71 microscope to galvo-galvo scanner. To make the alignment easier, an additional red laser was used. The laser propagates from the microscope objective toward the pinhole through the galvo-galvo scanner.

Careful alignment at this stage is critical because even a small angular error can significantly complicate subsequent alignment procedures. When the red laser enters the microscope objective, its image can be observed at the microscope camera port. The alignment beam is then directed toward the galvo-galvo scanner output port by using an alignment plate and physically moving either the galvo-galvo scanner or the microscope position. The beam position is then verified after the beamsplitter to ensure that the optical axis is properly directed toward the pinhole.

\subsection{Excitation laser to pinhole alignment}

There are two goals for this step: (1) to align the excitation laser onto the sample plane and (2) to ensure that the reflected or emitted light from the sample is efficiently coupled through the pinhole.

The alignment begins by steering the green excitation laser through the galvo-galvo scanner using the kinematic mirror in the excitation path and the beamsplitter. A sheet of white paper placed between the scanner and beamsplitter is used to compare the positions of the red alignment laser and the green excitation laser. The green beam is adjusted until it is concentric with the red beam. Once both beams overlap at this location, the remaining optical path is typically aligned automatically because both beams share the same optical axis.

After the beam has been aligned through the microscope, the lenses forming the excitation-path $4f$ relay are installed. The position of each lens is adjusted such that the beam passes through its center. The separation between the lenses must remain equal to the designed $4f$ distance to preserve proper imaging conditions.

The collection path is then optimized using a calibration microscope slide. A calibration target with a minimum feature size of \SI{0.01}{\milli\meter} is placed on the microscope stage, and the objective focus is adjusted until the grid pattern is sharply resolved. A white paper screen positioned above the slide can be used to visualize the illuminated region.

Fine adjustments of the focus are performed until the laser spot is confined to a single grid feature. The sample position is then adjusted in the $x-y$ plane so that a grid line passes through the center of the laser spot. Additional $z-$axis adjustments are made until the illuminated region appears dark, indicating that the excitation beam is reflected by the calibration grid rather than transmitted through it. As the sample is translated across the grid pattern, alternating bright and dark regions should be observed.

Once this condition is achieved, a power meter is placed after the pinhole. The kinematic mirror immediately before the pinhole and the axial position of the collection lens are adjusted to maximize the measured optical power. This procedure ensures that the reflected light from the sample is properly focused through the pinhole.

\subsection{Pinhole to the SPD alignment}

The final stage consists of coupling the fluorescence signal exiting the pinhole into the SPD. Ideally, the beam is relayed to the detector through the fluorescence-path $4f$ system. In practice, additional optics may be introduced to improve coupling efficiency by reducing the beam diameter and matching it to the acceptance conditions of the SPD fiber collimator.

For alignment, the optical fiber is temporarily disconnected from the SPD and connected to a calibrated optical power meter. The kinematic mirrors located after the pinhole are adjusted to optimize the beam position, while the axial positions of the coupling lenses are adjusted to match the beam size to the fiber input.

The alignment is considered optimized when the optical power coupled into the multimode fiber is maximized. In our setup, coupling efficiencies of approximately \SI{80}{\percent} are routinely achieved. After optimization, the optical fiber is reconnected to the SPD for fluorescence measurements.

\section{Standard Confocal Imaging Procedure}
\label{appendix:Standard Confocal Imaging Procedure}
\setcounter{figure}{0}
\setcounter{table}{0}

The following procedure was consistently used for image acquisition with the Olympus-based confocal microscope platform. This protocol ensures optimal PL signal collection from NV centers embedded in diamond particles.

First, the sample is mounted on the stage of the Olympus IX71 inverted microscope. The $z-$axis position is manually adjusted while monitoring real-time photon counts to bring the sample into approximate focus. Once the photon counts have been optimized, a broad scan is initiated. This generates an initial low-resolution image of the entire field of view.

From the acquired image, a bright spot of interest is selected, and a zoom scan is performed in the software to refine the spatial region. This zooming and scanning step can be repeated multiple times until a clearly defined region is identified. After determining the desired target area, the laser beam is manually positioned at the center of the selected bright spot, and the autofocus routine is executed. This triggers the piezo objective ($z-$axis) to perform a fine scan and automatically select the focal plane with the highest photon count.

To further maximize the signal, the user manually fine-tunes the microscope's mechanical $z-$stage and iteratively adjusts the $x/y$ positions of the kinematic mirrors of the excitation and collection paths. These minimal adjustments are repeated until the photon counts are optimized.

Finally, a high-resolution scan is performed over the region of interest to obtain a diffraction-limited image suitable for subsequent analysis.

\section{Troubleshooting}
\label{appendix:Troubleshooting}
\setcounter{figure}{0}
\setcounter{table}{0}

\subsection{SPD saturation}

\begin{figure}[h]
  \centering
  \includegraphics[width=\columnwidth]{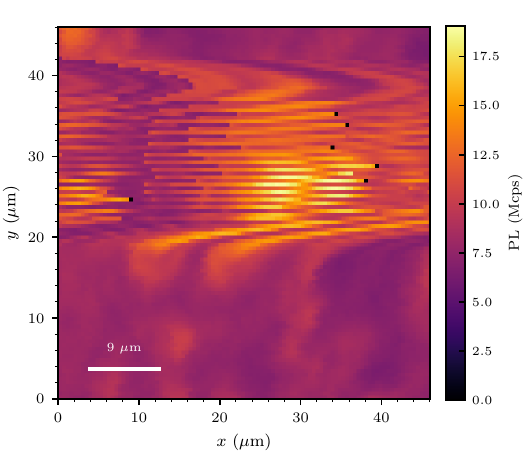}
  \caption{Confocal imaging artifacts when SPD saturation occurs.}
  \label{fig:Saturation}
\end{figure}

When the SPD becomes saturated distinct horizontal artifacts appear in the scan image as shown in Figure~\ref{fig:Saturation}. These artifacts manifest as bright horizontal lines across the image, corresponding to rows where the photon count exceeded the detector's linear response range. Saturation can also be confirmed by a visual alarm implemented in the real time acquisition window in the GUI software. The issue was mitigated by introducing neutral density filters as needed in front of the detector. For low-emission samples no ND filters were required due to significantly reduced PL intensity.

\subsection{Image edge artifacts}

\begin{figure}[h]
  \centering
  \includegraphics[width=\columnwidth]{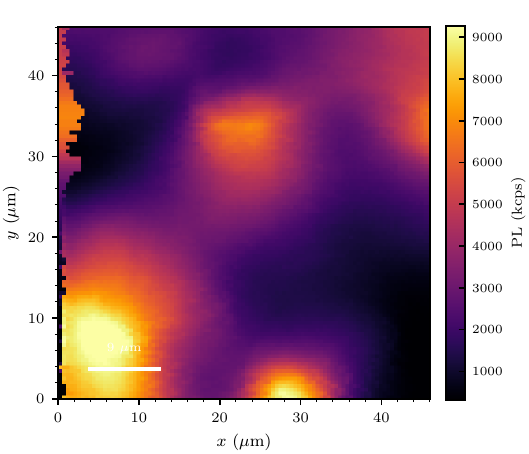}
  \caption{Artifacts at the left of the image caused by fast scans.}
  \label{fig:EdgeArtifacts}
\end{figure}

Blurring and non-uniform brightness near the left edges of scan images were consistently observed as observed in Figure~\ref{fig:EdgeArtifacts}. These artifacts were traced to rapid galvo mirror movements without sufficient stabilization time. Adding a short delay before each scan line effectively suppressed these edge artifacts and was implemented in the software. The delay time depends on the technical details of the galvo-galvo scanner.

\section{FPGA-based (Xilinx RFSoC 4x2 board) data acquisition and experiment control}
\label{appendix:RFSoC4x2}
\setcounter{figure}{0}
\setcounter{table}{0}

As an alternative to the proposed hardware configuration, the system was also tested using the Xilinx Radio Frequency System-on-Chip (RFSoC 4x2) board to acquire data during the imaging process and to control quantum sensing experiments via the QICK-DAWG framework~\cite{Riendeau2023}. A branch called \texttt{feature/RFSoC4x2} is available on our GitHub repository. 

The primary advantage of the RFSoC 4x2 board is a simpler hardware configuration: the microwave pulsed signal is synthesized directly on the board, in the same manner as the readout signal is read and analyzed in real-time. Pulse sequence programs and data analysis are provided through the QICK-DAWG python interface, which supports the collection and characterization of PL intensity, ODMR, $T_1$ relaxation time, Rabi oscillations, Ramsey, and Hahn echo sequences. 

\begin{figure}[h]
  \centering
  \includegraphics[width=1\columnwidth]{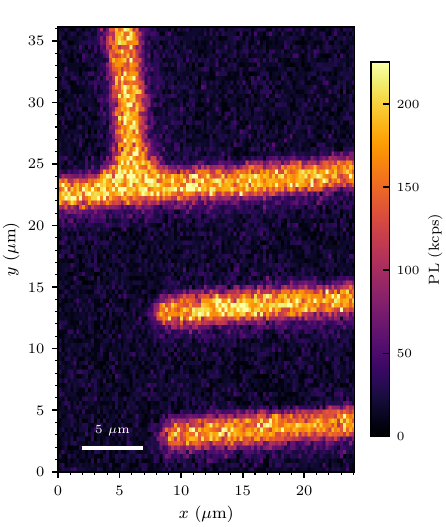}
  \caption{Confocal image acquired using the RFSoC4x2 FPGA board.}
  \label{fig:RFSoC4x2}
\end{figure}

However, in its current version the QICK-DAWG framework is not  optimized for confocal imaging, resulting in longer acquisition  times compared with the Time Tagger approach. For a $100 \times 100$  pixel image, the acquisition time using the RFSoC 4x2 board was  $\approx\SI{180}{\s}$, whereas the Time Tagger required only  $\approx\SI{20}{\s}$. An example image acquired using the RFSoC 4x2  is shown in Figure~\ref{fig:RFSoC4x2}.

A further limitation of the RFSoC 4x2 and QICK-DAWG combination arises in quantum experiments with a single NV center, where the  signal intensity is inherently low. In the current implementation, the maximum integration time per acquisition window is limited to $\SI{213}{\micro\s}$, a constraint imposed by the finite bit-depth  of the tProc internal counters in the QICK firmware (base firmware of QICK-DAWG)~\cite{Stefanazzi2022}. This  restriction has a significant impact on the signal-to-noise ratio (SNR) of single-NV measurements, which are fundamentally limited  by shot noise.

In photon-counting experiments, the shot noise arises from the Poissonian statistics of photon detection events. For $N$ detected photons, the shot noise standard deviation is $\sigma=\sqrt{N}$ yielding a contrast-weighted SNR that scales as

\begin{equation}
    \text{SNR} = \frac{C\sqrt{MN}}{\sqrt{1+C}}
    \label{eq:SNR}
\end{equation}

\noindent where $C$ is the measurement contrast, $N$ is the number of photons collected per repetition, and $M$ is the total number of repetitions. In our confocal system, using a $\SI{100}{\micro\m}$  pinhole, a typical single NV center yields a photon count rate of $\SIrange{10}{20}{kcps}$ in the bright state. Within the maximum integration window of $\SI{213}{\micro\s}$, this corresponds to only $N=2-4~\text{photons}$ per acquisition window. For a typical Single-NV ODMR  contrast of $C \approx 0.03$--$0.05$, achieving a target  $\text{SNR} \geq 10$ from Eq.~\ref{eq:SNR} requires the number  of repetitions to satisfy

\begin{equation}
    M \geq \frac{10^{2}(1+C)}{C^{2} N} \approx 
    10^{4}\text{ to } 6\times10^{4}
    \label{eq:repetitions}
\end{equation}

Substantially increasing the total acquisition time and  limiting the practical throughput of the system for single-NV  quantum sensing protocols such as ODMR, $T_1$, Rabi, Ramsey,  and Hahn echo sequences.

Despite these current limitations, active development is underway to optimize the QICK-DAWG framework for confocal imaging and  single-NV quantum sensing experiments using the RFSoC 4x2 board.  Efforts are focused on achieve imaging speeds  comparable to the Time Tagger approach. Full integration of the  RFSoC 4x2 into our confocal platform is particularly compelling  because it consolidates microwave pulse synthesis, readout, and  data acquisition into a single board, substantially reducing  hardware complexity relative to the configuration described in  Figure~\ref{fig:simplifiedschematic}. Furthermore, the significantly lower cost of  the RFSoC 4x2 board compared to dedicated hardware  lowers the barrier to entry for laboratories seeking to implement  NV-based quantum sensing and confocal imaging without access to expensive  instrumentation, broadening the accessibility of these techniques  to the wider research community.

\section{Live cell imaging using Zeiss LSM 900 super resolution confocal microscope}

\begin{figure}[h]
  \centering
  \includegraphics[width=1\columnwidth]{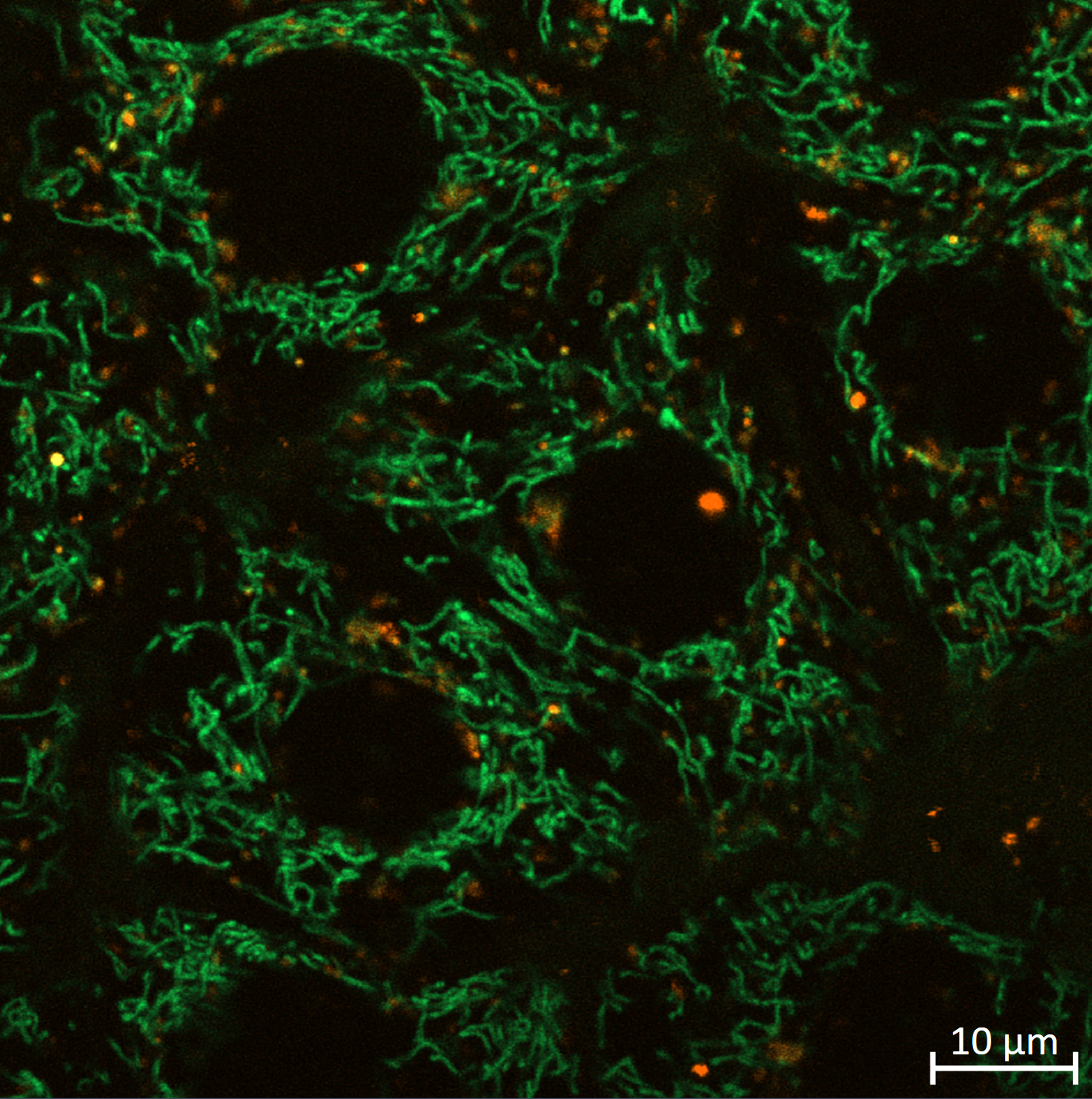}
  \caption{Dual-color imaging using Zeiss LSM 900 super resolution confocal microscope.}
  \label{fig:dualcolorzeiss}
\end{figure}

In order to have a comparison point for our Olympus-based confocal microscope for MagLOV cells with NVs, we used a Zeiss LSM 900 confocal microscope equipped with an Airyscan detector. A 63$\times$ oil-immersion objective with a NA of 1.4. to image MagLOV cells fluorescence. The sample was excited using a \SI{488}{\nano\meter} laser at \SI{1}{\percent} laser power, and emission was collected between \SIrange{500}{550}{\nano\meter}. The results are shown in Figure~\ref{fig:dualcolorzeiss} and the sample was the same as the one reported in this work, see Figure~\ref{fig:dualcolor}.

\section{Laser diode current driver}
\label{appendix:LaserDiodeCurrentDriver}

As an alternative to acousto-optic modulators (AOMs) for laser power modulation, we developed an electronic laser pulse driver capable of directly modulating the current of a laser diode. The design is based on the work of Sewani et al.~\cite{Sewani2020}, but incorporates several important improvements. First, the biasing network was redesigned to minimize transient ringing during pulse operation. Second, the entire circuit was migrated to a commercially manufacturable printed circuit board (PCB) design using EasyEDA and released as an open-source hardware project on GitHub (\url{https://github.com/burkelabuci/qlaser}).

The development of this driver was motivated by the limited availability and long procurement times of AOMs, which are commonly used to generate the laser pulses required for quantum-state initialization and readout. To avoid this bottleneck, we investigated direct current modulation of laser diodes, such as the Thorlabs LP520-SF15A operating at 520 nm.

\begin{figure*}[htbp]
  \centering
  \includegraphics[width=\textwidth]{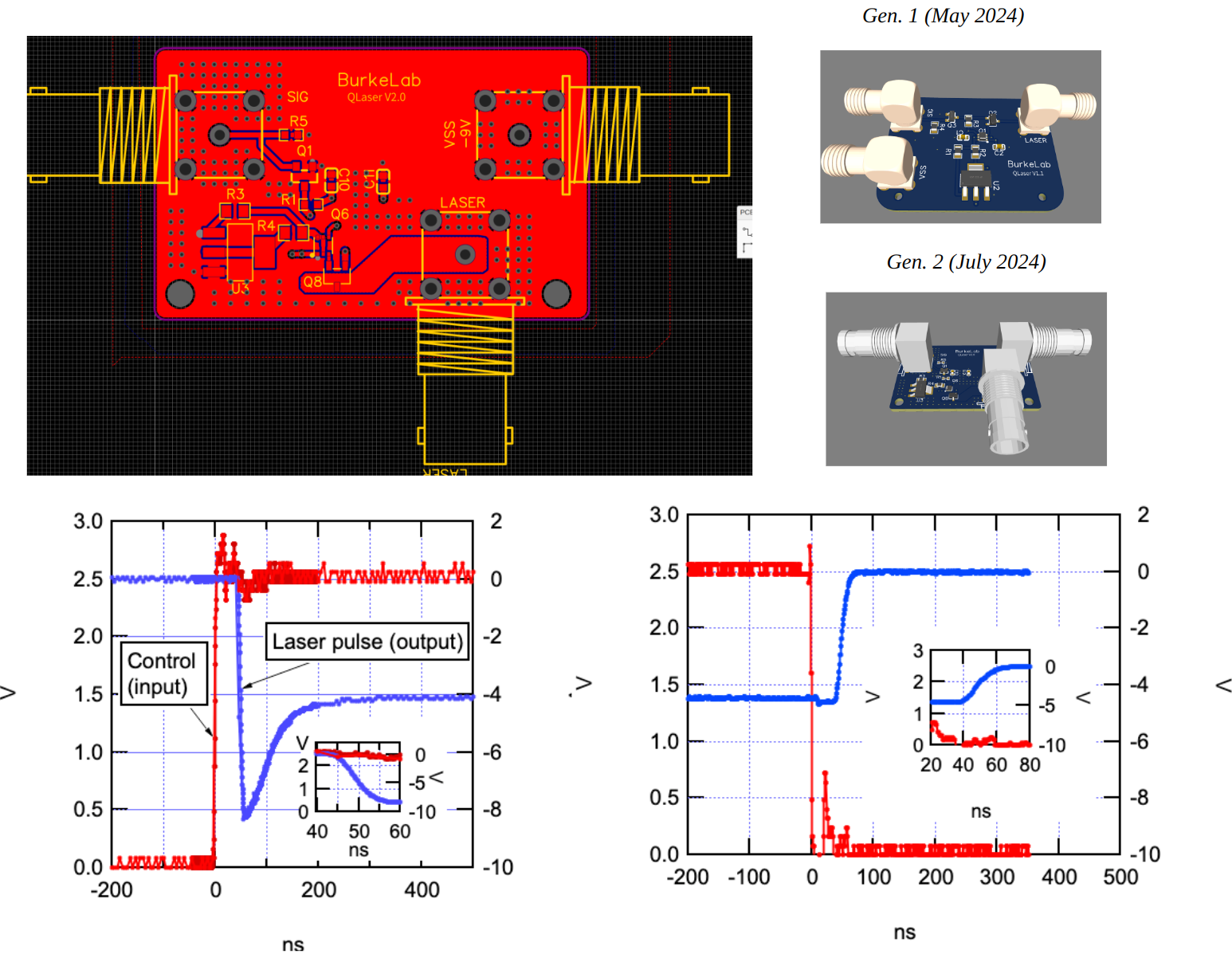}
  \caption{Laser pulse generator. The first generation had significant overshoot. The second generation is shown in the next figure.}
  \label{fig:LaserDriverCircuit}
\end{figure*}

\begin{figure}[htbp]
  \centering
  \includegraphics[width=1\columnwidth]{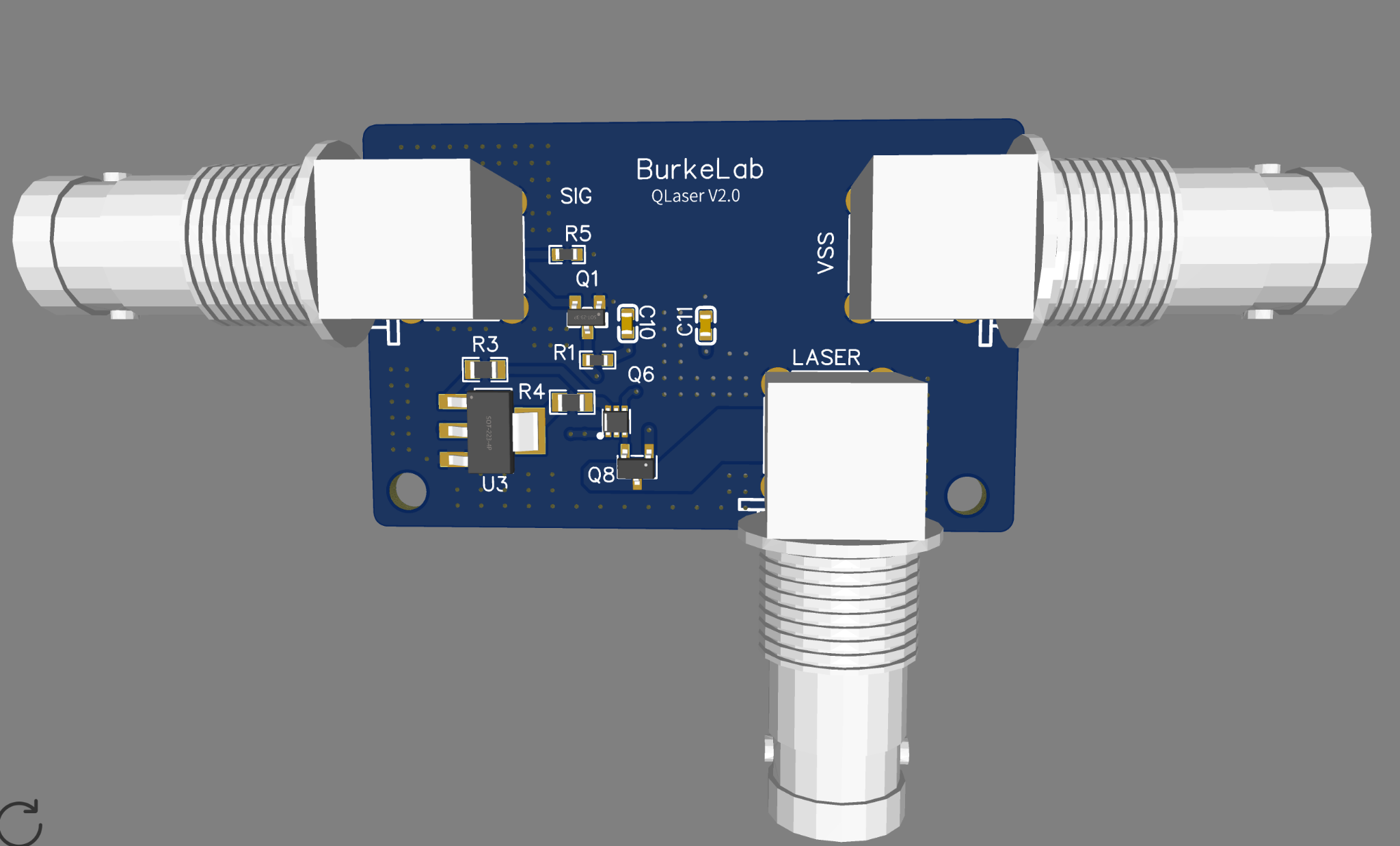}
  \caption{Laser pulse generator, version 2.0. This entire driver can be purchased fully assembled with no soldering required for around \$100 for 5 pieces including shipping.}
  \label{fig:LaserDriverBoard}
\end{figure}

The resulting pulse driver, shown in Figures~\ref{fig:LaserDriverCircuit} and \ref{fig:LaserDriverBoard}, provides a low-cost and readily accessible alternative to an AOM. The complete hardware design is open source and available at \url{https://github.com/burkelabuci/qlaser}. While the original design reported by Sewani et al.~\cite{Sewani2020} required manual assembly and hand soldering of individual components, our implementation was specifically optimized for commercial PCB assembly. All components, including connectors, can be factory assembled, allowing the board to be delivered fully populated and ready for use without additional fabrication effort.

\begin{figure*}[htbp]
  \centering
  \includegraphics[width=\linewidth]{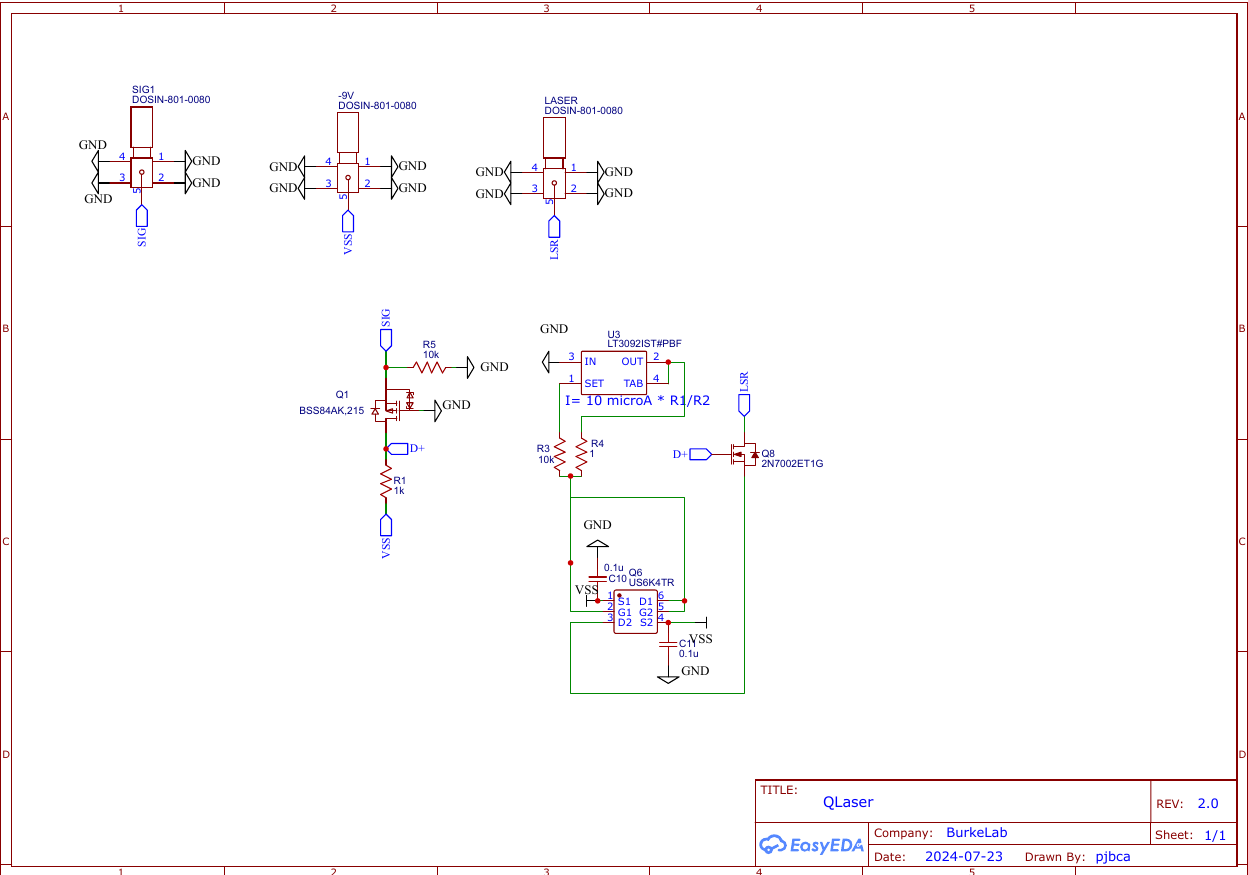}
  \caption{Laser driver circuit schematic.}
  \label{fig:LaserDriverCircuitSchematic}
\end{figure*}

In addition to simplifying manufacturing, the PCB layout incorporates several high-speed design improvements based on the authors' expertise in RF and microwave electronics. These modifications improve pulse integrity and facilitate reliable nanosecond-scale laser modulation. The total fabrication cost is approximately \$100 for a batch of five assembled boards, making the solution substantially less expensive than a typical AOM system, which can cost several thousand dollars. Consequently, this design provides an attractive alternative for laboratories that either cannot justify the cost of an AOM or face long procurement delays. The circuit schematic is shown in Figure~\ref{fig:LaserDriverCircuitSchematic}.

\begin{figure}[htbp]
  \centering
  \includegraphics[width=1\columnwidth]{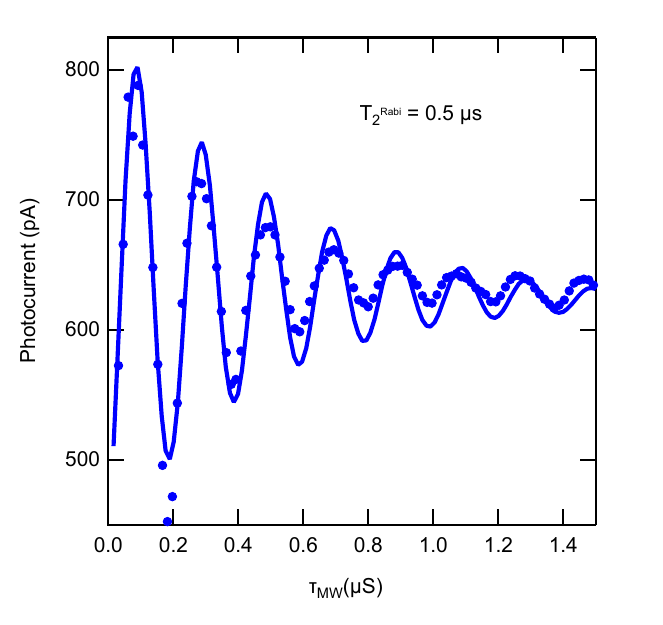}
  \caption{Rabi oscillations on ensembles of NV centers using electrically pulsed laser source.}
  \label{fig:RabiLayoutMethodsPaperV1}
\end{figure}

\begin{figure*}[htbp]
  \centering
  \includegraphics[width=0.9\linewidth]{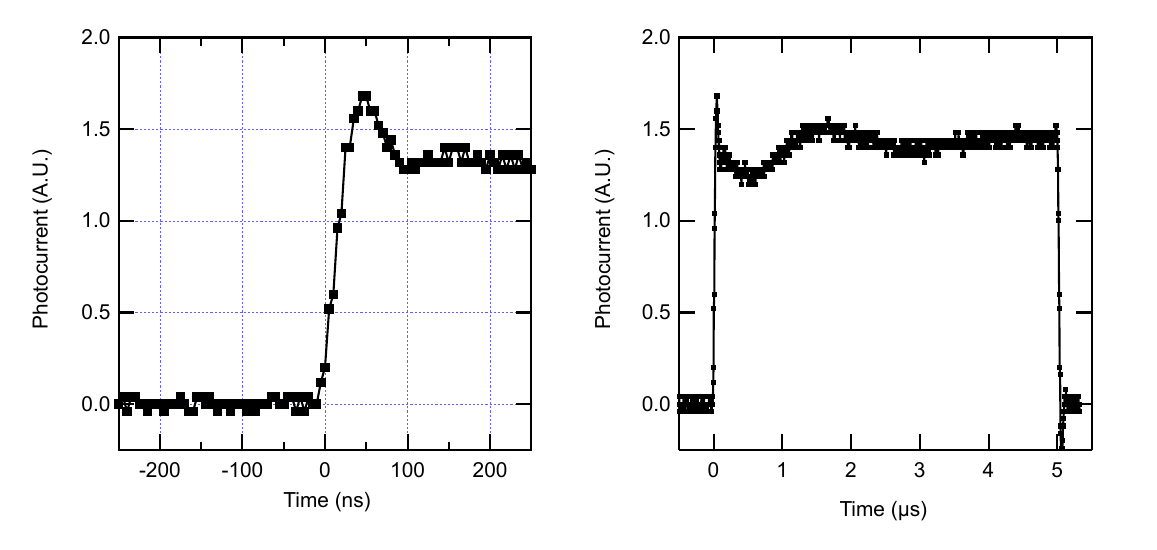}
  \caption{Photocurrent vs. time, showing \SI{25}{\nano\second} rise time for electrically driven laser.}
  \label{fig:RiseTime}
\end{figure*}

In order to benchmark the rise time, we measured the photocurrent from a pulsed laser, shown in Figure~\ref{fig:RiseTime}. The bias was adjusted down to \SI{6}{\volt} from the initial design of \SI{9}{\volt}, which significantly reduced the ringing, as seen in the figure. The rise time of the photocurrent (10-90\%) was \SI{25}{\nano\second}. This was sufficient to observe Rabi oscillations in ensembles of NV centers as shown in Figure~\ref{fig:RabiLayoutMethodsPaperV1}.

\section{Green Excitation Options: 532 nm Coherent Laser vs 520 nm Thorlabs Diode}
\label{appendix:laser-comparison}
\setcounter{figure}{0}
\setcounter{table}{0}

The excitation parts list (Table~\ref{tab:table1}) offers two green-laser configurations suited to different cost, pulsing, and dual-color requirements. Option~1 is built around a fiber-coupled Coherent Sapphire FP laser (\SI{532}{\nano\meter}, \SI{120}{\milli\watt}, CW). Option~2 replaces this source with a single-mode fiber-coupled Thorlabs LP520-SF15A laser diode (\SI{520}{\nano\meter}, \SI{15}{\milli\watt}), pulsed directly by the qlaser current driver described in Appendix~\ref{appendix:LaserDiodeCurrentDriver}. Both options share the same galvo-scanned $4f$ relay and detection path described in Section~\ref{sec:System Design}. The principal tradeoffs are summarized in Table~\ref{tab:laser-comparison}.

The first option is the conventional choice for pulsed NV experiments: a high-power CW laser gated by one or two fiber-coupled AOMs in series provides high extinction, nanosecond switching. In practice, however, AOMs presented significant obstacles during platform development. A fiber-coupled \SI{532}{\nano\meter} AOM (G\&H 1200AFP-AD-1.0) carried an eight-month lead time and was found to be non-functional upon delivery due to a design defect identified only after an extended warranty investigation. A subsequent dual-AOM free-space configuration using available laboratory units proved unreliable, with only one of three units performing within specification. The resulting optical path required a dedicated auxiliary table and introduced alignment complexity incompatible with the dual-color requirement of the biological imaging channel, since fiber-coupled AOMs optimized for \SI{532}{\nano\meter} cannot be repurposed for \SI{450}{\nano\meter} excitation.

These constraints motivated the adoption of the second approach direct current modulation of a fiber-pigtailed laser diode using the open-hardware qlaser driver (Appendix~\ref{appendix:LaserDiodeCurrentDriver}). This approach achieves a \SI{25}{\nano\second} rise time (10-90\%), which is sufficient for ODMR, Rabi, Ramsey, and $T_1$ protocols on, as demonstrated during previous experiments using ensamble NVs. The \SI{520}{\nano\meter} diode is also compatible with the dual-color configuration: the same dichroic combiner that merges the \SI{450}{\nano\meter} MagLOV channel accepts \SI{520}{\nano\meter} without modification, enabling both excitation wavelengths to share the full $4f$ relay and galvo-scanning path.

\section{Prior Art and Comparison}
\label{sec:Prior Art and Comparison}

Our instrument sits at the intersection of two distinct literatures -- open / teaching-lab NV microscopy and time-resolved fluorescence microscopy of biological radical pairs -- and differs from every previously published platform in each. In this section, we summarize the relevant prior art and make those differences explicit. 

\subsection{Solid-state NV-center microscopes}

Four previously reported platforms define the current landscape of NV-center instrumentation that is, at least in principle, reproducible outside a dedicated quantum-optics group. Zhang and co-workers~\cite{Zhang2018} described a minimal, continuous-wave ODMR platform based on a Mini-Circuits voltage-controlled oscillator and a simple photodiode readout, with total build cost between $\sim$\$1k (magnetometer variant) and $\sim$\$3k (fluorescence-microscope variant); ensemble nanodiamond was used, and no pulsed experiments were performed. Sewani and co-workers~\cite{Sewani2020} extended the teaching-lab concept to ensemble pulsed NV quantum sensing, delivering the complete CW-ODMR $\rightarrow$ Rabi $\rightarrow$ Hahn-echo $\rightarrow$ CPMG cascade at a build cost of $\sim$\$18.6k on $\langle 111\rangle$ HPHT diamond, with a PCB loop-gap-resonator microwave antenna. Misonou and co-workers~\cite{Misonou2020} presented a compact tabletop single-NV magnetometer explicitly designed to be reproducible ``by nonspecialists'', reporting Rabi, Ramsey ($T_2^* = \SI{0.50}{\micro\second}$), Hahn echo ($T_2 = \SI{364}{\micro\second}$), and single-ensemble proton NMR ($T_2 = \SI{16.1}{\micro\second}$, $d_\text{NV} = \SI{6.26}{\nano\meter}$) on a Keio/RIKEN-built instrument. Yuan, de Leon and co-workers~\cite{Yuan2024} reported an instructional single-NV confocal microscope at a build cost of \$68k--\$100k, demonstrating $g^{(2)}$(0) < 0.5 single-NV verification, \SI{38}{\nano\second} Rabi, $^{15}$N hyperfine Ramsey, and Hahn-echo $T_2$ at 23 G external magnetic field. Taken together, references~\cite{Zhang2018,Sewani2020,Misonou2020,Yuan2024} form a clean cost/capability ladder from $\sim$\$1k CW-only ODMR to $\sim$\$100k single-NV coherent control.

Our platform differs from each of references~\cite{Zhang2018,Sewani2020,Misonou2020,Yuan2024} in five concrete respects, none of which is merely a cost optimization.

\begin{itemize}
    \item \textbf{Commercial inverted microscope body with stationary-sample geometry.} References~\cite{Zhang2018,Sewani2020,Misonou2020,Yuan2024} are all built on cage-optics frames; the sample is moved on a stage (Misonou) or the objective on a piezo with a fixed diamond (Yuan). We instead build on an Olympus IX71 inverted microscope body, which is the standard chassis of mammalian cell biology, and scan the excitation beam by a galvo-galvo pair through $4f$ relays while holding the sample stationary. This decouples the biological sample environment from the scanning mechanism and makes the instrument immediately compatible with live-cell culture dishes, microfluidic manifolds, patch-clamp electrodes, and objective-side optical tweezers.
    \item \textbf{Two complementary control-electronics paths, benchmarked on one optical bench.} References~\cite{Misonou2020,Yuan2024} use a Tektronix AWG7102 + Anritsu MG3700A combination ($\gg$ \$20k) or an equivalent arbitrary-waveform / vector-signal-generator pair, and references~\cite{Sewani2020} relies on a SpinCore PulseBlaster ESR Pro 250 plus an external lock-in. We evaluate two alternatives on the same optical train -- an open-source RFSoC stack running QICK-DAWG~\cite{Riendeau2023} on a Real Digital RFSoC4x2 board ($\approx$ \$2k), and a dedicated hardware time tagger -- and show that they occupy two distinct, complementary operating regimes of NV sensing. This comparative benchmarking is itself a contribution of the present work and is expanded in the subsection below.
    \item \textbf{Open-hardware pulsed laser driver (qlaser) in place of fiber-coupled acousto-optic modulators.} High-extinction optical gating for pulsed NV experiments is conventionally implemented with fiber-coupled AOMs such as the Gooch \& Housego Fiber-Q used in reference~\cite{Misonou2020}. These modules cost of order \$10k each, two are typically required in series to achieve a combined extinction ratio sufficient for single-NV $T_1$ measurements, and, at the time of writing, lead times of six months or more are routine due to persistent supply-chain constraints. We replace the AOM pair with qlaser~\cite{PeterJBurke2024}, an open-hardware pulsed current source we have released under the GNU GPL v3 license at github.com/burkelabuci/qlaser. The board is a modified derivative of the pulsed laser driver described in Sewani et al.~\cite{Sewani2020}; five fully assembled boards can be ordered from a standard prototyping fabricator for $\sim$\$100 total, i.e. of order \$10--\$20 per drive channel. Direct current-mode pulsing of the laser diode yields nanosecond rise and fall times sufficient for ODMR, Rabi, Ramsey and $T_1$ protocols without any moving optical element.    
    \item \textbf{Demonstrated compatibility with live, respiring mitochondria.} None of the NV microscopes in references~\cite{Zhang2018,Sewani2020,Misonou2020,Yuan2024} has been applied to a living biological sample. The combination of a stationary-sample inverted geometry, and a dual-excitation optical path at \SI{520}{\nano\meter} (NV) and \SI{450}{\nano\meter} (flavin / MagLOV / EYFP) makes the present instrument the first NV confocal platform whose optics, mechanics and electronics have been jointly validated for imaging vital, metabolically active mitochondria. The live-mitochondrial imaging demonstrated in Section~\ref{sec:experiments} is, to our knowledge, the first of its kind on a single-NV-capable microscope.        
    \item \textbf{A single optical path designed to connect solid-state and biological qubit protocols.} References~\cite{Zhang2018,Sewani2020,Misonou2020,Yuan2024} are monochromatic NV instruments. Our $4f$-relay architecture combines a \SI{520}{\nano\meter} NV excitation path with a reconfigurable \SI{450}{\nano\meter} blue-excitation path for flavin / MagLOV / EYFP-class biological fluorescence measurements, collects emission through the same confocal detection framework, and gates detection with a single-photon timing module. In the present work, the experimentally demonstrated spin measurements are NV ODMR and $T_1$ relaxation. The same optical, timing, microwave, and RF architecture is designed to support future Rabi, Ramsey, $T_2$, and time-resolved RYDMR measurements, but those experiments are not claimed as completed results here.    
\end{itemize}

\subsection{Biological-qubit fluorescence microscopes}

The instrumentation most directly comparable on the biological-qubit side is the fluorescence-microscopy platform of Ikeya and Woodward~\cite{Ikeya2021,NoboruIkeya2026}. Their 2021 PNAS instrument~\cite{Ikeya2021} is a custom, upright fluorescence microscope based on a 100$\times$/NA 1.49 oil objective, a \SI{450}{\nano\meter} CW diode laser at $\sim$\SI{0.5}{\kilo\watt\per\centi\meter\squared}, an sCMOS camera, and a GMW 5204 vector electromagnet. It established that endogenous HeLa cell autofluorescence responds to static magnetic fields up to \SI{25}{\milli\tesla} with a triplet-born flavin SCRP signature ($B_{1/2}$ = 18.0 $\pm$ 0.5 mT, saturation MFE 3.7\%). Their 2026 JACS platform~\cite{NoboruIkeya2026} adds two identical \SI{450}{\nano\meter} nanosecond pulsed lasers combined on a multimode fiber, a rapidly switched magnetic field coil with < 30 ns rise/fall time, and LabVIEW / Raspberry-Pi-Pico orchestration, and introduces the pump--probe (PP) and pump--field--probe (PFP) techniques, resolving radical-pair lifetimes down to $\sim$50 ns in subcellular ($\sim$1--4 pL) volumes.

Our instrument differs from references~\cite{Ikeya2021,NoboruIkeya2026} in four concrete respects.

\begin{itemize}
    \item \textbf{Microwave and radio-frequency spin control on the biological-qubit channel.} References~\cite{Ikeya2021,NoboruIkeya2026} measure magnetic-field effects only: a static or a rapidly switched magnetic field alters the SCRP singlet--triplet branching, and the fluorescence change is recorded. They do not deliver a coherent microwave or RF drive to the biological qubit and therefore cannot perform RYDMR, Rabi or Ramsey experiments on the SCRP itself. Our platform integrates a \SIrange{0.5}{1.0}{\giga\hertz} RF drive channel, matched to the $\sim$604 MHz S--T splitting of engineered MagLOV2 SCRPs~\cite{Abrahams2026}, together with the \SI{2.87}{\giga\hertz} NV microwave chain. We can therefore perform true pulsed RYDMR and, in principle, coherent S--T control on biological qubits -- something not possible on references~\cite{Ikeya2021,NoboruIkeya2026}.
    \item \textbf{Solid-state qubit capability on the same bench.} References~\cite{Ikeya2021,NoboruIkeya2026} have no NV channel: no green excitation, no \SI{2.87}{\giga\hertz} microwave resonator, no pulsed-ODMR readout. The biological-qubit and solid-state-qubit communities have so far run on separate instruments. The present platform is, to our knowledge, the first reported microscope that natively supports both protocols at the same diffraction-limited spot, making it the first plausible experimental venue for hybrid NV--biological-qubit measurements.
    \item \textbf{Inverted, stationary-sample geometry for live-cell imaging. }References~\cite{Ikeya2021,NoboruIkeya2026} are upright and use a 100$\times$/NA 1.49 oil objective pressed against a coverglass from above. Live-cell culture chambers, incubation inserts and patch-clamp manipulators are designed for inverted microscopes; our Olympus IX71 chassis and stationary-sample galvo-galvo geometry therefore enables classes of experiment that are not accessible to the Woodward platform.
    \item \textbf{Super-resolution compatible optics and sub-diffraction quantum sensing. }The Woodward PFP instrument reports an $\sim$8 $\mu$m detection spot and therefore averages over tens of mitochondria. Our galvo-galvo $4f$-relay optics support single-point confocal addressing at the diffraction limit and, in combination with the super-resolution MagLOV ODMR approach we have recently demonstrated in live mammalian cells in CW mode~\cite{Burke2026}, provides a natural route to super-resolved, time-resolved RYDMR on mitochondrion-targeted biological qubits. The pulsed, RF-addressable variant of reference~\cite{Burke2026} is one of the explicit motivations for the present instrument.
\end{itemize}

\section{Coordinated experimental program with the Ikeya--Woodward platform}
\label{appendix:Coordinated experimental program with the Ikeya--Woodward platform}

The most productive relationship between the Ikeya--Woodward platform and the present Olympus-based pulsed quantum microscope is not competitive but sequential. We propose, and our platform is designed to support, the following coordinated program.

\begin{enumerate}
    \item[(i)] Replicate the IW26 validation suite, including FMN, FAD, FMN/TrpH, and FAD/TrpH at pH~2.3, on our platform using PP and PFP pulse programs implemented on QICK-DAWG~\cite{Riendeau2023} and qlaser~\cite{PeterJBurke2024}. A successful cross-validation would place the two instruments on a common methodological footing.

    \item[(ii)] Extend from incoherent magnetic-field gating to coherent RF drive at $\sim$~\SI{604}{\mega\hertz} on MagLOV2~\cite{Alkahtani2026} and related engineered flavoproteins. This is the extension IW26 explicitly identifies as future work, and our RF chain is designed for this operating regime.

    \item[(iii)] Introduce NV magnetometry as an in-situ calibration method for the rapidly switched magnetic field. In this mode, the transient field profile measured by a surface-proximal NV center can be compared directly with the IW26 Hall-sensor calibration.

    \item[(iv)] Demonstrate pulsed RYDMR on mitochondrially targeted biological qubits, following the super-resolution continuous-wave approach of our earlier work~\cite{Burke2026}, while using the NV channel to benchmark the local magnetic-field environment in the same optical voxel.

    \item[(v)] Ultimately, prepare and measure an entangled state shared between an NV center in a surface-proximal nanodiamond and a biological qubit in the adjacent cell. This experiment is accessible, in principle, only on the class of platform reported here, but it is best understood as the endpoint of the IW26 $\rightarrow$ MagLOV $\rightarrow$ entangled-sensor trajectory that IW26 itself begins.
\end{enumerate}

\subsection{Summary of novelty}

The novelty of the present work is therefore not a single new completed measurement, but the integration of microscope-level capabilities that have each existed separately in the literature and whose combination opens a new experimental regime.

\begin{itemize}
  \item[(i)] We unify, in one inverted-microscope body, the green / \SI{2.87}{\giga\hertz} NV channel of references~\cite{Sewani2020,Misonou2020,Yuan2024}, the optical geometry required for blue / nanosecond-pulsed biological-qubit excitation in references~\cite{Ikeya2021,NoboruIkeya2026}, and the RF-frequency range relevant to MagLOV-class biological qubits in references~\cite{Abrahams2026,Antill2026,Burd2026}. In the present work, the experimentally demonstrated NV spin measurements are ODMR and $T_1$ relaxation. The same architecture is designed to support future $T_2$, Ramsey, Hahn-echo, and time-resolved RYDMR measurements after the corresponding pulse-sequence validation.
  \item[(ii)] We present and benchmark two complementary control-electronics paths on the same optical bench: an open-source RFSoC stack running QICK-DAWG~\cite{Riendeau2023} on a Real Digital RFSoC4x2 board for multi-NV ensemble operation, and a dedicated hardware time tagger for single-NV operation. 
  \item[(iii)] We adopt a stationary-sample galvo-galvo geometry on a standard Olympus IX71 body, eliminating sample-motion artifacts associated with sample-scanning implementations in references~\cite{Sewani2020,Misonou2020,Yuan2024} and making the instrument natively compatible with inverted biological-imaging formats.
  \item[(iv)] We demonstrate biological fluorescence compatibility on a single-NV-capable optical platform by imaging TMRE-stained and MagLOV-expressing HeLa cells. This result does not yet constitute biological-qubit RYDMR, but it establishes that the same microscope framework used for NV confocal imaging, ODMR, and $T_1$ measurement can be reconfigured for genetically encoded biological fluorescence samples through limited optical changes.
  \item[(v)] The architecture is furthermore a natural optical front end for the noise-suppressed NV-pair gradiometer protocol we have recently proposed for biological operation~\cite{Alkahtani2026}. The present data establish the required confocal imaging, ODMR, $T_1$ readout, stationary-sample scanning, and biological fluorescence compatibility, while two-qubit entanglement-enhanced sensing remains a future experimental target inspired by references~\cite{Rovny2025,Zhou2025,Le2025}.
\end{itemize}

\section{Parts List}
\label{appendix:parts list}
\setcounter{figure}{0}
\setcounter{table}{0}

Here we organized parts list with different options. You can build your setup according to your budget

\begin{table*}[h]
  \caption{\label{tab:laser-comparison}Comparison of green excitation options.}
  \begin{ruledtabular}
  \begin{tabular}{lll}
  Property & Option 1 (Coherent Sapphire) & Option 2 (Thorlabs LP520) \\
  \hline
  Wavelength                  & \SI{532}{\nano\meter}            & \SI{520}{\nano\meter}                        \\
  Output power                & 120~mW                           & 15~mW                                        \\
  Fiber coupling              & Single-mode, factory             & Single-mode, factory                         \\
  Pulsing method              & External AOM required            & Direct current modulation (qlaser)           \\
  Rise/fall time              & ${<}$10~ns (AOM-limited)         & ${\sim}$25~ns (qlaser-limited)               \\
  Dual-color compatibility    & No (AOM is 532~nm-specific)      & Yes (dichroic combination at input)          \\
  Approximate cost            & ${\sim}$\$5k (laser) $+$ ${\sim}$\$10k (AOM) & ${\sim}$\$800 (laser) $+$ ${\sim}$\$20 (driver board) \\
  \end{tabular}
  \end{ruledtabular}
\end{table*}

\begin{table*}[t]
\caption{\label{tab:table1}List of parts for the 4f system of the laser excitation setup.}
\begin{ruledtabular}
\begin{tabular}{lllll}
Item & Vendor & Model & Description & Qty \\
\hline
\multicolumn{5}{l}{\textbf{Laser Option 1:}} \\[2pt]
Green Laser & Coherent & Sapphire FP & 532 nm, 120 mW, fiber-coupled & 1 \\[6pt]

\multicolumn{5}{l}{\textbf{Laser Option 2:}} \\[2pt]
Green Laser Diode & Thorlabs & LP520-SF15A & 520 nm, 15 mW, SM Fiber-Pigtailed Laser Diode, FC/PC & 1 \\[6pt]
Blue Laser Diode & Thorlabs & LPF450-SF25 & 450 nm, 25 mW, SM Fiber-Pigtailed Laser Diode, FC/PC & 1 \\[6pt]

\multicolumn{5}{l}{\textbf{AOM 1:}} \\[2pt]
AOM & ISOMET & 1205C-2-790 & 2 mm aperture, 3.63mm/$\mu$s acoustic velocity& 1 \\[6pt]
AOM Driver & ISOMET & 222R-2-81 & 50 $\Omega$, 80 MHz, 1.6 W & 1 \\[6pt]

\multicolumn{5}{l}{\textbf{AOM 2:}} \\[2pt]
AOM+Driver & Brimrose & TEM-200-25-20-532-2FP & Fiber coupled, 532 nm, 200 MHz, 1W & 1 \\[6pt]

\multicolumn{5}{l}{\textbf{Other Optics:}} \\[2pt]

Fiber Adaptor Plate & Thorlabs & SM1FCA2 & FC/APC Fiber Adaptor & 3 \\
Mounted Aspheric Lens & Thorlabs & C280TMD-A & Collimate light exiting fiber/diode & 3 \\
Lens Cell Adaptor & Thorlabs & S1TM09 & Aspheric Lens holder & 3 \\
Kinematic Mount & Thorlabs & KC1-T & For aspheric lens/fiber adapter plate & 3 \\
Lens Tube & Thorlabs & SM1L10 & For aspheric lens focus/lens hood & 3 \\
30 mm Cage Plate & Thorlabs & CP45 & Removable cage plate & 3 \\
SM1 Cage Plate & Thorlabs & CP33 & For cage rods and mounting lenses & 3 \\
Cage Assembly Rod & Thorlabs & R4-P4 & 4" Long, Ø6 mm, 4 Pack & 3 \\
Z-axis Translation Mount & Thorlabs & SM1ZA & Fine tuning distance between lenses & 3 \\
Plano-Convex Lens & Thorlabs & LA1251-AB & f=100 mm & 5 \\
Plano-Concave Lens & Thorlabs & LC4513-AB & f=-7.5 cm & 3 \\
Bandpass Filter & Thorlabs & FLH532-4 & 532 nm bandpass filter & 1 \\
Beamsplitter & Thorlabs & BSN10 & Ø1", 10:90, 400-700 nm & 1 \\
Kinematic Mount & Thorlabs & KM100 & For reflection mirror & 6 \\
Centering Plate & Thorlabs & KCP1 & For kinetic mount & 6 \\
Centering Plate & Thorlabs & KCP1 & For kinetic mount & 6 \\
Lens Mount & Thorlabs & LMR1 & Fixed lens mount & 4 \\
Iris & Thorlabs & CP20D & For alignment & 1 \\
Iris & Thorlabs & ID25 & For alignment & 3 \\
Optical Post & Thorlabs & TR1 & 1.0'' post & 4 \\
Optical Post & Thorlabs & TR1.5 & 1.5'' post & 22 \\
Post Holder & Thorlabs & PH1.0 & 1'' holder & 4 \\
Post Holder & Thorlabs & PH1.5 & 1.5'' holder & 22 \\
Clamping Fork & Thorlabs & CF125 & For beamsplitter & 8 \\
Pedestal Base Adapter & Thorlabs & BE1 & For beamsplitter & 8 \\
Reflection Mirror & Thorlabs & PF10-03-P01 & Ø1'' Protected Silver Mirror & 5 \\
Spacer & Thorlabs & RSxM & Height adjustment  & as needed \\
\end{tabular}
\end{ruledtabular}
\end{table*}

\begin{table*}[t]
\caption{\label{tab:table2}List of parts for the 4f system of the fluorescence collection setup.}
\begin{ruledtabular}
\begin{tabular}{lllll}
Item & Vendor & Model & Description & Qty \\
\hline
Achromatic Doublet Lens & Thorlabs & AC254-150-AB & f=150 mm & 1 \\
Plano-Convex Lens & Thorlabs & LA1251-AB & f=100 mm, N-BK7 & 2 \\
Z-axis Translation Mount & Thorlabs & SM1ZA & Fine tuning between lenses & 1 \\
Pinhole & Thorlabs & P50K1 & 50 µm, SM1-threaded, mounted & 1 \\
Longpass Filter & Thorlabs & FELH0600 & 600 nm cut-on & 1 \\
ND Filter Wheel & Thorlabs & FW1AND & 5 ND filters included & 1 \\
Fiber Adaptor Plate & Thorlabs & SM1FC2 & FC/PC Fiber Adaptor, For SPD & 1 \\
Mounted Aspheric Lens & Thorlabs & C280TMD-A & Collimate light exiting fiber/diode & 1 \\
Lens Cell Adaptor & Thorlabs & S1TM09 & Aspheric Lens holder & 1 \\
Kinematic Mount & Thorlabs & KC1-T & For aspheric lens/fiber adapter plate & 1 \\
Lens Tube & Thorlabs & SM1L10 & For aspheric lens focus/lens hood & 1 \\
Fiber Patch Cable & Thorlabs & M42L01 & FC/PC multimode fiber & 1 \\
SPD & Excelitas & SPCM-AQRH-10-FC & 1500 cps dark, 22 ns dead time & 1 \\
Kinematic Mount & Thorlabs & KM100 & For mirrors & 3 \\
Reflection Mirror & Thorlabs & PF10-03-P01 & Ø1'' Protected Silver Mirror & 3 \\
Centering Plate & Thorlabs & KCP1 & For kinetic mount & 3 \\
Optical Post & Thorlabs & TR1.5 & 1.5'' post & 13 \\
Post Holder & Thorlabs & PH1.5 & 1.5'' holder & 13 \\
Cage Cube & Thorlabs & CM1-DCH & For beamsplitter mounting & 1 \\
SM1 Cage Plate & Thorlabs & CP33 & For cage rods & 3 \\
Cage Assembly Rod & Thorlabs & R4-P4 & 4" Long, Ø6 mm, 4 Pack & 1 \\
Spacer & Thorlabs & RSxM & Height adjustment  & as needed \\
\end{tabular}
\end{ruledtabular}
\end{table*}

\begin{table*}[t]
\caption{\label{tab:table3}List of parts for the spectroscopy setup.}
\begin{ruledtabular}
\begin{tabular}{lllll}
Item & Vendor & Model & Description & Qty \\
\hline
Beam sampler & Thorlabs & BSF10-A & 1" beam sampler & 1 \\
Kinematic Mount & Thorlabs & KM100C & For beamsplitter plate & 1 \\
Diffraction Grating & Thorlabs & GT50-03 & Visible transmission grating & 1 \\
Plano-Convex Lens & Thorlabs & LA1251-AB & f=100 mm & 2 \\
Monochrome CCD & Player One & Poseidon-M Pro (IMX571) & Cooled camera for spectra & 1 \\
Kinematic Mount & Thorlabs & KM100 & For mirrors & 3 \\
Reflection Mirror & Thorlabs & PF10-03-P01 & Ø1'' Protected Silver Mirror & 3 \\
Centering Plate & Thorlabs & KCP1 & For kinetic mount & 3 \\
Optical Post & Thorlabs & TR1.5 & 1.5'' post & 6 \\
Post Holder & Thorlabs & PH1.5 & 1.5'' holder & 6 \\
Spacer & Thorlabs & RSxM & Height adjustment  & as needed \\
\end{tabular}
\end{ruledtabular}
\end{table*}

\begin{table*}[t]
\caption{\label{tab:table4}List of parts for the 4f system of the nanoscale positioning in the experiment.}
\begin{ruledtabular}
\begin{tabular}{lllll}
Item & Vendor & Model & Description & Qty \\
\hline
Galvo-Galvo Scanner & Thorlabs & LSKGG4 & XY scan, ±15°, 20 µrad resolution & 1 \\
Kinematic Mount & Thorlabs & KM100 & For mirrors & 2 \\
Reflection Mirror & Thorlabs & PF10-03-P01 & Ø1'' Protected Silver Mirror & 2 \\
Centering Plate & Thorlabs & KCP1 & For kinetic mount & 2 \\
SM1 Cage Plate & Thorlabs & CP33 & For cage rods and mounting lenses & 3 \\
Cage Assembly Rod & Thorlabs & R4-P4 & 4" Long, Ø6 mm, 4 Pack & 1 \\
Plano-Convex Lens & Thorlabs & LA1251-AB & f=100 mm & 1 \\
Iris & Thorlabs & CP20D & For alignment & 2 \\
Objective Piezo & Thorlabs & PFM450E & Z scan, 3 nm resolution, 450 µm range & 1 \\
Microscope Objective & Olympus & UPLSAPO40X2 & 40x/0.95, WD 0.18 mm & 1 \\
\end{tabular}
\end{ruledtabular}
\end{table*}

\begin{table*}[t]
\caption{\label{tab:table5}List of parts for fluorescence imaging and laser monitoring.}
\begin{ruledtabular}
\begin{tabular}{lllll}
Item & Vendor & Model & Description & Qty \\
\hline
Microscope objective & Zeiss & 420792-9800-000 & Alpha Plan-Apo 100x/1.46Oil DIC M27 & 1 \\
Notch Filter & Thorlabs & NF514-17 & 514 nm, FWHM = 17 nm & 2 \\
CMOS Camera & Thorlabs & CS165CU & Sample Imaging & 1 \\
Power Meter & Thorlabs & PM100D & 100 pW - 200 W range & 1 \\
Power Sensor & Thorlabs & S120C & 400-1100 nm, 50 nW - 50 mW & 1 \\
Reflection Mirror & Thorlabs & PF10-03-P01 & Ø1'' Protected Silver Mirror & as needed \\
Optical Post & Thorlabs & TR1.5 & 1.5'' optical post & as needed \\
Post Holder & Thorlabs & PH1.5 & 1.5'' holder & as needed \\
Kinematic Mount & Thorlabs & KM100 & For mirror/collimator & as needed \\
\end{tabular}
\end{ruledtabular}
\end{table*}

\begin{table*}[t]
\caption{\label{tab:table6}List of electronics used in the experiment.}
\begin{ruledtabular}
\begin{tabular}{lllll}
Item & Vendor & Model & Description & Qty \\
\hline
Pulse Streamer & Swabian & PS 8.2 & 8 digital/2 analog outputs & 1 \\
Time Tagger & Swabian & Time Tagger 20 & 8 inputs, 34 ps jitter & 1 \\
DAQ & NI & USB-6453 & 32 AI channels, 20-bit & 1 \\
Microwave Amplifier & Mini-Circuits & ZHL-16W-43-S+ & 1.8-4 GHz, 16 W & 1 \\
IQ Modulator & TI & TRF37t05EVM & Microwave drive/modulation & 1 \\
Circulator & Mini-Circuits & PE83CR005 & 2-4 GHz & 1 \\
Directional Coupler & Mini-Circuits & ZDFC-20-33-S+ & 20-3000 MHz, 20.5 dB & 1 \\
Step Attenuator & -- & -- & 1 dB - 10 dB steps & 1 \\
Spectrum Analyzer & Rigol & DSA832E & 9 kHz-3.2 GHz & 1 \\
High Power Terminator & Fairview & ST6S-10 & SMA terminator, used on circulator port & 1 \\
Oscilloscope & Rigol & DS1054 & 4 ch, 50 MHz, 1 GSa/s & 1 \\
MW Cables & Thorlabs & SMMxx & 50 $\Omega$, DC-26 GHz & several \\

SMA 50 $\Omega$ Terminator & -- & -- & For port termination/impedance matching & several \\
SMA-F / SMA-M Adapters & -- & -- & Gender adapters for SMA connectors & several \\
\end{tabular}
\end{ruledtabular}
\end{table*}

\end{document}